\documentclass[twocolumn,superscriptaddress,floatfix,prl,showpacs,nolongbibliography]{revtex4-2}
\usepackage{graphicx}
\usepackage{amsmath}
\usepackage{amssymb}
\usepackage{color}
\usepackage{braket}
\usepackage{dsfont}
\usepackage{lineno}
\usepackage{float}

\usepackage[resetlabels]{multibib}
\newcites{Supp}{suppmaterial}
\begin{document}
\title{Signatures of Dissipation Driven Quantum Phase Transition in Rabi Model}

\author{G. De Filippis}
\affiliation{SPIN-CNR and Dip. di Fisica E. Pancini - Universit\`a di Napoli Federico II - I-80126 Napoli, Italy}
\affiliation{INFN, Sezione di Napoli - Complesso Universitario di Monte S. Angelo - I-80126 Napoli, Italy}

\author{A. de Candia}
\affiliation{SPIN-CNR and Dip. di Fisica E. Pancini - Universit\`a di Napoli Federico II - I-80126 Napoli, Italy}
\affiliation{INFN, Sezione di Napoli - Complesso Universitario di Monte S. Angelo - I-80126 Napoli, Italy}

\author{G. Di Bello}
\affiliation{Dip. di Fisica E. Pancini - Universit\`a di Napoli Federico II -\
 I-80126 Napoli, Italy}

\author{C. A. Perroni}
\affiliation{SPIN-CNR and Dip. di Fisica E. Pancini - Universit\`a di Napoli Federico II - I-80126 Napoli, Italy}

\author{L.~M.~Cangemi}
\affiliation{Department of Chemistry, Bar-Ilan University, Ramat-Gan 52900, Israel}

\author{A. Nocera}
\affiliation{Department of Physics and Astronomy and Stewart Blusson Quantum Matter Institute, University of British Columbia, Vancouver, B.C., Canada, V6T 1Z1}

\author{M. Sassetti}
\affiliation{Dipartimento di Fisica, Universit\`a di Genova, I-16146 Genova, Italy}
\affiliation{SPIN-CNR, I-16146 Genova, Italy}

\author{R. Fazio}
\affiliation{SPIN-CNR and Dip. di Fisica E. Pancini - Universit\`a di Napoli Federico II - I-80126 Napoli, Italy}
\affiliation{ICTP, Strada Costiera 11, I-34151 Trieste, Italy}
\affiliation{NEST, Istituto Nanoscienze-CNR, I-56126 Pisa, Italy}

\author{V. Cataudella}
\affiliation{SPIN-CNR and Dip. di Fisica E. Pancini - Universit\`a di Napoli Federico II - I-80126 Napoli, Italy}
\affiliation{INFN, Sezione di Napoli - Complesso Universitario di Monte S. Angelo - I-80126 Napoli, Italy} 

\begin{abstract}
  By using worldline Monte Carlo technique, matrix product state and a variational approach \`a la Feynman, we investigate the equilibrium properties and relaxation features of the dissipative quantum Rabi model, where a two level system is coupled to a linear harmonic oscillator embedded in a viscous fluid. We show that, in the Ohmic regime, a Beretzinski-Kosterlitz-Thouless quantum phase transition occurs by varying the coupling strength between the two level system and the oscillator. This is a non perturbative result, occurring even for extremely low dissipation magnitude. By using state-of-the-art theoretical methods, we unveil the features of the relaxation towards the thermodynamic equilibrium, pointing out the signatures of quantum phase transition both in the time and frequency domains. We prove that, for low and moderate values of the dissipation, the quantum phase transition occurs in the deep strong coupling regime. We propose to realize this model by coupling a flux qubit and a damped LC oscillator.
\end{abstract}
\maketitle

In 1936 Rabi introduced a model describing the simplest class of light-matter interaction, i.e. the dipolar coupling between a two-level quantum system (qubit) and a classical monochromatic radiation field (unidimensional harmonic oscillator) \cite{rabi}. In its quantum version \cite{cummings,zueco1,braak}, i.e. the so-called quantum Rabi model, the radiation is specified by a quantized single-mode field. In general, the interaction between an atom and the electromagnetic field inside a cavity allows us to get not only a deep understanding of the light-matter interaction, but it also plays a significant role in different quantum technologies, including lasers and many quantum computing architectures \cite{review,science} like ultrafast gates \cite{romero}, quantum error correcting codes \cite{kiaw}, remote entanglement generation \cite{felicetti}, cold atoms and trapped ions \cite{review1}. Recently, the coherent coupling of a single photon mode and a superconducting charge qubit has been extensively studied both from theoretical \cite{rossatto,zueco,grifoni,grifoni1,larson} and experimental points of view \cite{1,2,3,4,nature}. Nowadays, the realization of strong, ultrastrong and deep strong coupling \cite{wendin,anote} between artificial atoms and cavities is possible, for instance by inductively coupling a flux qubit and an LC oscillator via Josephson junctions \cite{nature}. Indeed, an important feature of the flux qubit is its strong anharmonicity: the two lowest energy levels are well isolated from the higher levels. In the most interesting regime, the deep strong coupling one, where the coupling strength becomes as large as the atomic and cavity frequencies, the energy eigenstates of the qubit-resonator system are highly entangled. On the other hand, one of the central problems is the full understanding of all the physical properties of such quantum systems when the interaction with environmental degrees of freedom, inducing decoherence and dissipation, is explicitly taken into account. Specifically, the questions we want to address in the present letter are: does the dissipative quantum Rabi model exhibit a quantum phase transition (QPT) and what is its signature in linear response measurements?

In the literature the existence of a QPT has been addressed in the Dicke model \cite{dicke,dicke1,new,new1} and the resistively shunted Josephson junction \cite{murani,j1,j2,j3,j4}. In the former case, describing a collection of N two-level atoms interacting with a single bosonic mode via a dipole interaction, it has been proved that the system undergoes a transition from quasi-integrability to quantum chaos, and that this transition is caused by the precursors of the QPT, occurring when $N\rightarrow\infty$. In the latter case, where a Josephson junction and its capacitor, analogous to a massive particle in a washboard potential, are coupled to a bath of harmonic oscillators that provides viscous damping, the existence of a QPT has given rise to a long-standing controversy \cite{murani}. Indeed, the absence of QPT, in the parameter regime predicted theoretically, has been reported \cite{murani}. In a simpler case, the spin-boson model, where a system with only two energy levels is coupled to an Ohmic environment, QPT existence has been well established \cite{weiss}. Indeed, by increasing the interaction between the qubit and the bath, QPT occurs.

Recently it has been shown that the Rabi Hamiltonian exhibits a QPT despite consisting only of a single-mode cavity field and a two-level atom \cite{plenio1,ashhab}. It appears when the cavity frequency $\omega_0$, in units of the qubit gap $\Delta$, tends to zero, i.e. QPT takes place in the classical limit for the harmonic oscillator. In particular, it has been proved that: 1) the number of spins in the Dicke model and the ratio $\Delta / \omega_0$ in the Rabi model play an identical role; 2) the open Rabi model exhibits a mean field second-order dissipative phase transition \cite{plenio2}. These predictions have been experimentally observed \cite{plenio3}.

In this letter we show that in the fully quantum limit, i.e. $\omega_0 / \Delta \ne 0$, the dissipative Rabi model exhibits another and completely different QPT: by increasing the qubit-resonator interaction a Beretzinski-Kosterlitz-Thouless (BKT) QPT occurs. In particular, we prove that this is a not perturbative result. Indeed, QPT takes place for any fixed, but nonvanishing, value of the coupling between the cavity and the bath. Firstly, by using worldline Monte Carlo (WLMC) method \cite{giulio1,giulio2,bnote}, based on the path integrals, and a variational approach \`a la Feynman \cite{Feynman,giulio1,giulio2,bnote}, we investigate the equilibrium properties of the dissipative quantum Rabi model. We prove that, in the Ohmic regime, a BKT QPT occurs by varying the coupling strength between the two level system and the oscillator, even for extremely low dissipation magnitude. In particular, by indicating with $\alpha_{cav}$ the strength of the coupling between the cavity and the bosonic bath, we show that QPT sets in when $4 g^2 \alpha_{cav} \simeq \omega_0^2$, i.e., for low and moderate values of the dissipation, $\alpha_{cav} \lesssim 0.25$, QPT occurs in the deep strong coupling regime that, nowadays, can be experimentally reached. Furthermore, by using matrix product state simulations (MPS) \cite{bnote,dmrg1,dmrg2, dmrg3, fishman, chinalex}, and combining the Mori formalism \cite{mori1,bnote} and a variational approach \`a la Feynman, we investigate also the relaxation processes towards the thermodynamic equilibrium. They allow us to identify the signatures of the QPT both in the time and frequency domains, establishing a relation between the order parameter and a typical linear response measurement like the magnetic susceptibility. 

{\it The Model.} The Hamiltonian is written as:
\begin{equation}\label{eq:definitionH}
  H=H_{Q-O} + H_{I},
\end{equation}
where: 1) $H_{Q-O}=-\frac{\Delta}{2}\sigma_{x}+\omega_0 a^\dagger a+ g \sigma_z (a+a^\dagger)$ describes the qubit-oscillator system, $\Delta$ being the tunneling matrix element, $a$ ($a^\dagger$) standing for the annihilation (creation) operator for the bosonic field with frequency $\omega_0$, and $g$ representing the strength of the coupling; 2) $H_{I}=\sum_{i=1}^N \left[ \frac{p_i^2}{2 M_i} + \frac{k_i}{2} (x-x_i)^2 \right]$ describes the environmental degrees of freedom and their interaction with the resonator. The bath is represented as a collection of harmonic oscillators with frequencies $\omega_i^2=\frac{k_i}{M_i}$, and coordinates and momenta given by $x_i$ and $p_i$ respectively; furthermore $x$ denotes the position operator of the resonator with mass $m$: $x=\sqrt{\frac{1}{2m\omega_0}}(a+a^\dagger)$. Units are such that $\hbar=k_B=1$. We emphasize that, in Eq.(\ref{eq:definitionH}), $\sigma_x$ and $\sigma_z$ are Pauli matrices with eigenvalues $1$ and $-1$. The dissipative environment is modeled as a strictly Ohmic bath with spectral density: $J(\omega)=\sum_{i=1}^N \frac{k_i \omega_i}{2m\omega_0}\delta(\omega-\omega_i)=\alpha_{cav}\omega \theta(\omega_c-\omega)$. Here the adimensional parameter $\alpha_{cav}$ measures the strength of the coupling and $\omega_c$ is a cutoff frequency. By means of the unitary transformation that diagonalizes the Hamiltonian of the cavity and its environment, the model can be mapped \cite{grifoni1,zueco,bnote} to the Hamiltonian of a single two level system, with gap $\Delta$, interacting, through $\sigma_z$ operator, with a structured bosonic bath. The effective spectral density function, $J_{eff}(\omega)=\sum_{i=1}^{N+1} l_i^2 \delta(\omega-\tilde{\omega}_i)$, is given by: 
\begin{equation} 
  J_{eff}(\omega)=\frac{2 g^2 \omega_0^2\alpha_{cav}\omega \theta(\omega_c-\omega)}{(\omega^2-\omega_0^2-h(\omega))^2+(\pi \alpha_{cav} \omega_0 \omega)^2},
  \label{spectralfunction}
\end{equation}
$\tilde{\omega}_i$ being the frequencies of the $N+1$ bosonic normal modes stemming from the diagonalization of the cavity-environment Hamiltonian, $l_i$ the couplings with $\sigma_z$, and $h(\omega)=\alpha_{cav} \omega_0 \omega \log \left [ \frac{\omega_c +\omega}{\omega_c - \omega}\right ]$. We emphasize that $J_{eff}(\omega)$ features a Lorentzian peak at the oscillator frequency $\omega_0$ with width $\pi \alpha_{cav} \omega_0$, and, at low frequencies, $\omega \ll \omega_0$, exhibits an Ohmic behavior, $J_{eff}(\omega) \simeq \frac{\alpha_{eff}}{2} \omega$, with $\alpha_{eff}=4 g^2 \alpha_{cav} / \omega_0^2$. In the following we set $\alpha_{cav}=0.2$, $\omega_0=0.75 \Delta$, and $\omega_c=10 \Delta$.

{\it QPT Evidences at the Thermodynamic Equilibrium.} We investigate the equilibrium properties by using two different approaches. The first one is a variational approach \`a la Feynman at finite temperature \cite{giulio1,giulio2,bnote}. The other one is WLMC method, based on the path integrals \cite{giulio1,giulio2,bnote}. Here the elimination of the structured bath degrees of freedom leads to an effective Euclidean action \cite{bulla_1,weiss,bnote,feynmannew}:
\begin{equation}\label{eq:eqS}
  S=\frac{1}{2}\int_0^{\beta} d\tau \int_0^{\beta} d\tau^{\prime} \sigma_{z}(\tau) K_{eff}(\tau-\tau^{\prime}) \sigma_{z}(\tau^{\prime}),
\end{equation}
where $\beta=1/T$ ($T$ is the system temperature), and the kernel is expressed in terms of the spectral density $J_{eff}(\omega)$ and the bath propagator: $K_{eff}(\tau)=\int_0^{\infty} d\omega J_{eff}(\omega) \frac{ \cosh \left[ \omega \left( \frac{\beta}{2}-\tau \right)\right]}{\sinh \left( \frac{\beta \omega}{2} \right)}$. The problem turns out to be equivalent to a classical system of spin variables distributed on a chain with length $\beta$, and ferromagnetically interacting with each other with strength $K_{eff}(\tau-\tau^{\prime})$ ($\tau$ and $\tau^{\prime}$ label the spins on the chain). The functional integral is done with Poissonian measure adopting a cluster algorithm \cite{rieger,bulla_1}, based on the Swendsen-Wang approach \cite{swang,bnote}. In particular, if $\omega_0$ is kept constant and $\beta\rightarrow\infty$, $K_{eff}$ has the asymptotic behavior: $K_{eff}(\tau)=\frac{\alpha_{eff}}{2 \tau^2}$. We will prove that it determines the onset of a BKT QPT.

\begin{figure}[H]
  \includegraphics[width=1.01\columnwidth]{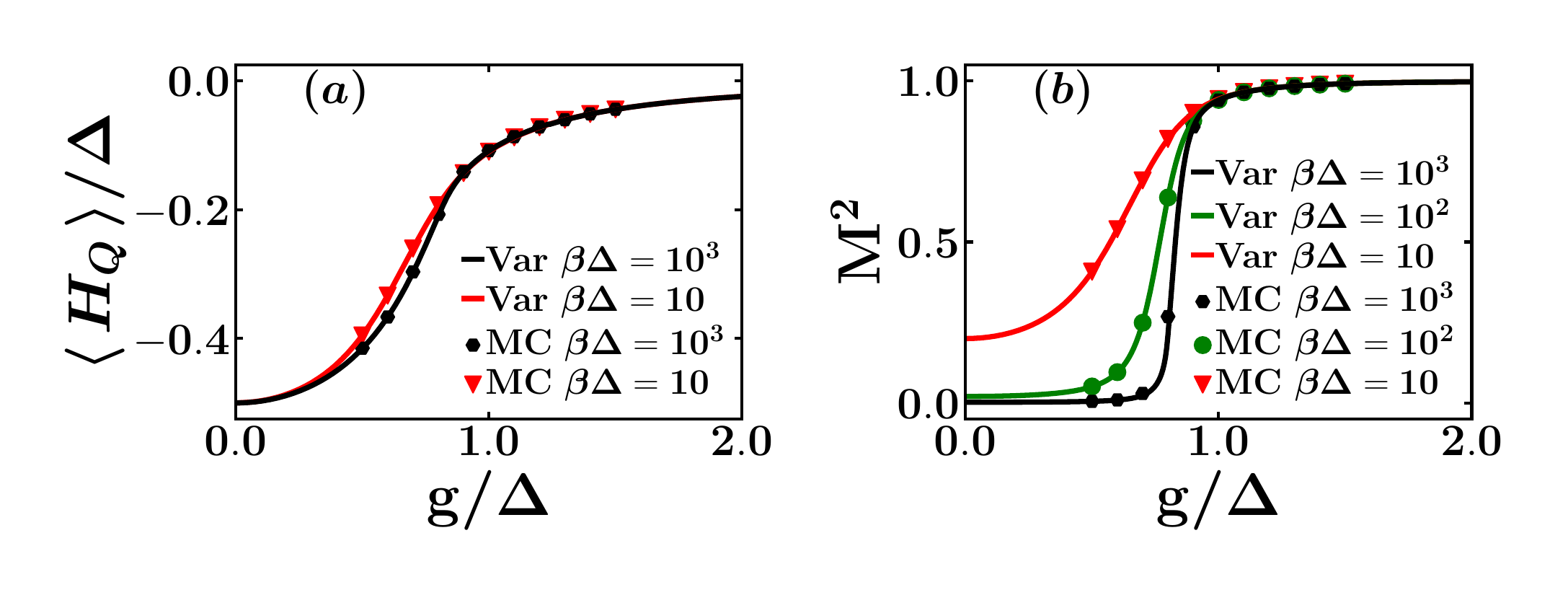}
  \caption{\label{fig:1} (color online)
    $\langle H_{Q} \rangle /\Delta$ (a), and $M^2$ (b) vs $g/\Delta$ at different temperatures: comparison between WLMC method and variational approach (MC and Var in the figure).}
\end{figure}

In Fig.~\ref{fig:1} we plot $\langle H_{Q} \rangle$, in units of $\Delta$, with $H_Q=-\frac{\Delta}{2}\sigma_{x}$, i.e. the two-level system Hamiltonian, and qubit squared magnetization, $M^2=\frac{1}{\beta}\int_{0}^{\beta} d\tau \langle \sigma_z(\tau) \sigma_z(0)\rangle$, as a function of $g/\Delta$, for different temperatures, from $T=10^{-1} \Delta$ to $T=10^{-3} \Delta$. The plots point out the successful agreement between the two proposed approaches. As expected, by increasing $g/\Delta$, $\langle H_{Q} \rangle$ increases, indicating a progressive reduction of the effective tunnelling. Interestingly, we emphasize that $\langle H_{Q} \rangle$ is always different from zero, even for extremely large values of $g/\Delta$, and slightly depends on the temperature.

\begin{figure}[H]
  \includegraphics[width=1.01\columnwidth]{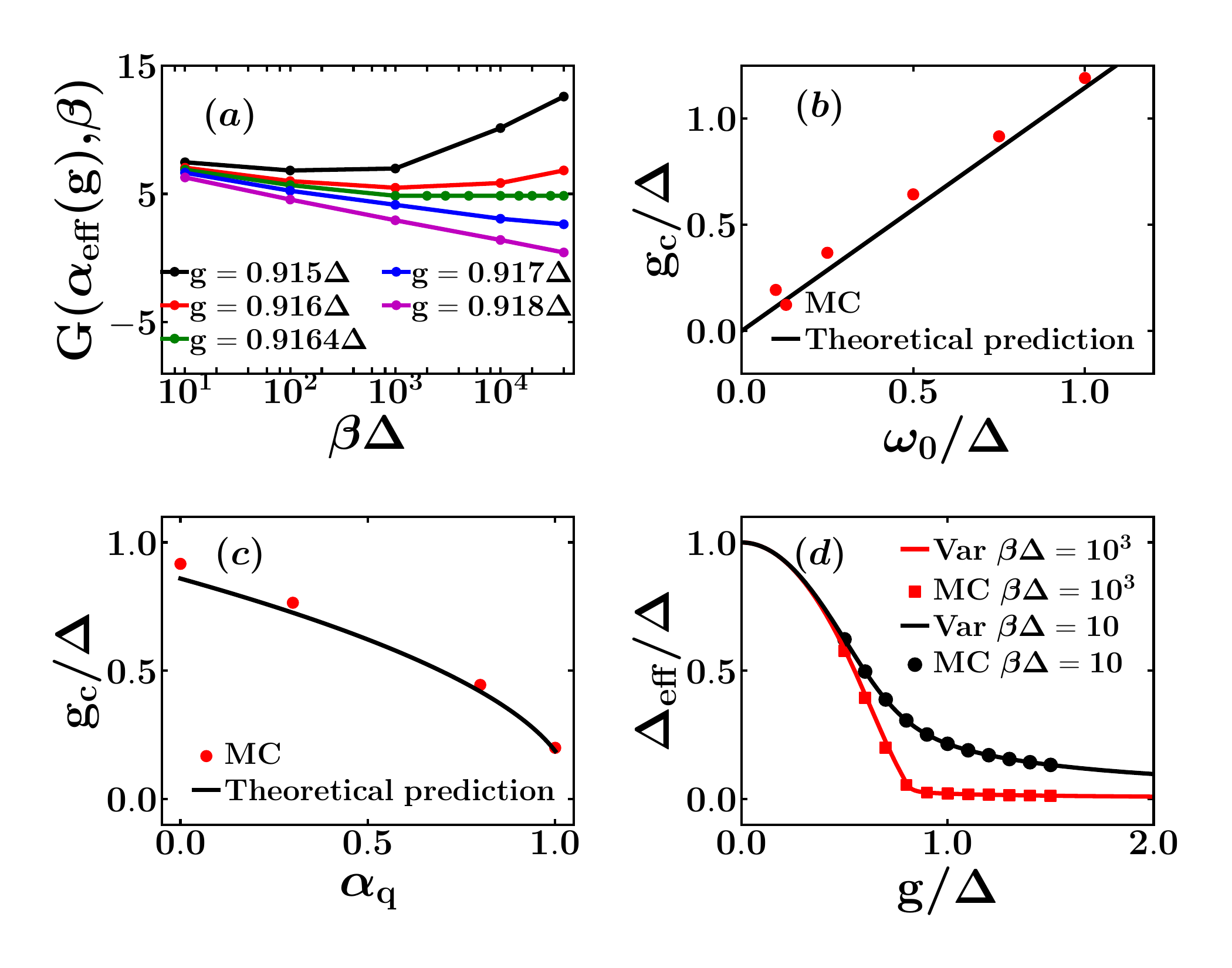}
  \caption{\label{fig:2} (color online)
    a) The function $G$ vs $\beta \Delta$ at $g \simeq g_c$ by using WLMC technique; $g_c/\Delta$ vs $\omega_0/\Delta$ (b) and $\alpha_q$ (c): comparison between WLMC method and an effective theory based only on the low frequency contribution of the spectral density ($\omega_0/\Delta=0.75$ in (c)); d) the qubit effective gap at two different temperatures: comparison between WLMC technique and variational approach (MC and Var in the figure).}
\end{figure}

On the other hand, $M^2$ increases from $0$ to about $1$, in a steeper and steeper way by lowering $T$, signaling an incipient QPT, that is a BKT QPT. Indeed, in a BKT transition, the quantity $M^2$ should exhibit a discontinuity at a critical value of $g/\Delta$ and $T=0$ \cite{Kosterlitz_1973,kosterlitz1}. In order to get a precise estimation of the critical value of the coupling, i.e. $g_c$, and then critical value of $\alpha_{eff}$, i.e. $\alpha_c$, we adapt the approach suggested by Minnhagen et al. in the framework of the XY model \cite{minnhagen_1,minnhagen_2,minnnew}. In the present context, the roles of the chirality and the lattice size are played by squared magnetization and inverse temperature $\beta$, respectively. Defining the scaled order parameter $\Psi(\alpha_{eff},\beta)=\alpha_{eff} M^2$, the BKT theory predicts for large values of $\beta$, i.e. asymptotically: $\frac{\Psi(\alpha_c,\beta)}{\Psi_c}=1+\frac{1}{2(\ln\beta-\ln\beta_0)}$, where $\beta_0$ is the only fitting parameter and $\Psi_c=\Psi(\alpha_c,\beta\rightarrow\infty)$ is the universal jump that is expected to be equal to one. In this scenario, the function $G(\alpha_{eff},\beta)=\frac{1}{\Psi(\alpha_{eff},\beta)-1}-2\ln\beta$ should not show any dependence on $\beta$ at $\alpha_{eff}=\alpha_c$. In Fig.~\ref{fig:2}a we plot the function $G(\alpha_{eff},\beta)$, as a function of $\beta$, for different values of $g/\Delta$, and then $\alpha_{eff}$. The plots clearly show that there is a value of $\alpha_{eff}$ such that $G$ is independent on $\beta$ asymptotically. This determines $\alpha_c$. In the presence of a purely Ohmic bath, i.e. $J(\omega)=\frac{\alpha}{2}\omega \Theta(\omega_c-\omega)$, the critical value of $\alpha$ is about $1.05$ at $\omega_c=10 \Delta$ \cite{giulio1}. In Fig.~\ref{fig:2}b we plot $g_c/\Delta$ vs $\omega_0/\Delta$ compared with that obtained by taking into account only the low-frequency contribution of the spectral density function, i.e. by imposing $\alpha_c=4 g_c^2 \alpha_{cav} / \omega_0^2=1.05$. The successful agreement clearly shows that QPT is driven by the asymptotic behavior of the spectral density, i.e. by the long range interaction of the mapped spin system that decays as $1/\tau^2$. In order to furtherly corroborate this observation, we take into account also the direct influence of the environment on the qubit, i.e. we add another contribution in the spectral density function of Equation (\ref{spectralfunction}): $\frac{\alpha_q}{2} \omega \Theta(\omega_c-\omega)$. It stems from the interaction between an Ohmic bath and the qubit through the operator $\sigma_z$. In Fig.~\ref{fig:2}c we compare $g_c/\Delta$ vs $\alpha_q$, computed by means of MC technique at $\omega_0/\Delta=0.75$, with that obtained by retaining only the low-frequency contribution in the bath spectral density, i.e. by imposing $\alpha_q+\alpha_{eff}=1.05$. The plot, also in this case, points out the robustness of the previously discussed hypothesis.
It is worth noting that the equation determining the QPT onset, i.e. $\alpha_c=4 g_c^2 \alpha_{cav} / \omega_0^2=1.05$, proves that, for $\alpha_{cav} \lesssim 0.25$, the quantum transition occurs in the deep strong coupling regime. We also emphasize that, within the BKT QPT scenario, i.e. $\omega_0$ is finite and $\beta \rightarrow \infty$, $g_c$ is proportional to $\omega_0$. On the other hand, when $\omega_0 \rightarrow 0$ and $\beta \rightarrow \infty$, with $\omega_0 \beta \rightarrow 0$, the kernel in Eq.(\ref{eq:eqS}) is independent of $\tau$, so that a mean field transition occurs \cite{bnote,plenio1,ashhab}. It is controlled by the adimensional parameter $\lambda=\frac{g^2}{\omega_0 \Delta}$ with $\lambda_c=\frac{1}{4}$, i.e. $g_c \propto \sqrt{\omega_0}$. In the supplemental material we also investigate the physical consequences on the resonator of the BKT QPT occurrence. Starting from the resonator Green function, $D(\tau)$, relative to resonator position operator $x$, we find an exact relation between $X^2=\frac{1}{\beta}\int_{0}^{\beta} d\tau \langle x(\tau)x(0)\rangle$ and $M^2$. We prove that both these physical quantities exhibit a discontinuity in the BKT QPT, whereas they increase linearly with $\lambda -\lambda_c$ in the mean field transition \cite{bnote,tnote}.  

{\it QPT Evidences from Relaxation Function and Magnetic Susceptibility.}
Let's suppose that the system at $t=-\infty$ is at the thermal equilibrium. The response of the system to a perturbation, adiabatically applied from $t=-\infty$ and cut off at $t=0$, can be calculated within the Mori formalism and the linear response theory \cite{kubo}. In particular, in the presence of a small magnetic field $h$ along $z$ axis, $\forall t \geq 0$ the most important physical quantity is the qubit relaxation function $\Sigma_{z}(t)=\frac{\langle \sigma_z(t) \rangle}{\langle \sigma_z(0) \rangle}$ (calculated in the absence of $h$, being $t \geq 0$). Within the Mori formalism \cite{mori1,bnote}, where the inner product between two operators is defined by $(A,B)= \frac {1}{\beta} \int_{0}^{\beta} \left\langle e^{sH} A^{\dagger} e^{-sH} B \right\rangle ds$, it is possible to prove that $\Sigma_{z}(t)=\frac{(\sigma_z(0),\sigma_z(t))}{(\sigma_z(0),\sigma_z(0))}$.

\begin{figure}[H]
  \includegraphics[width=1.01\columnwidth]{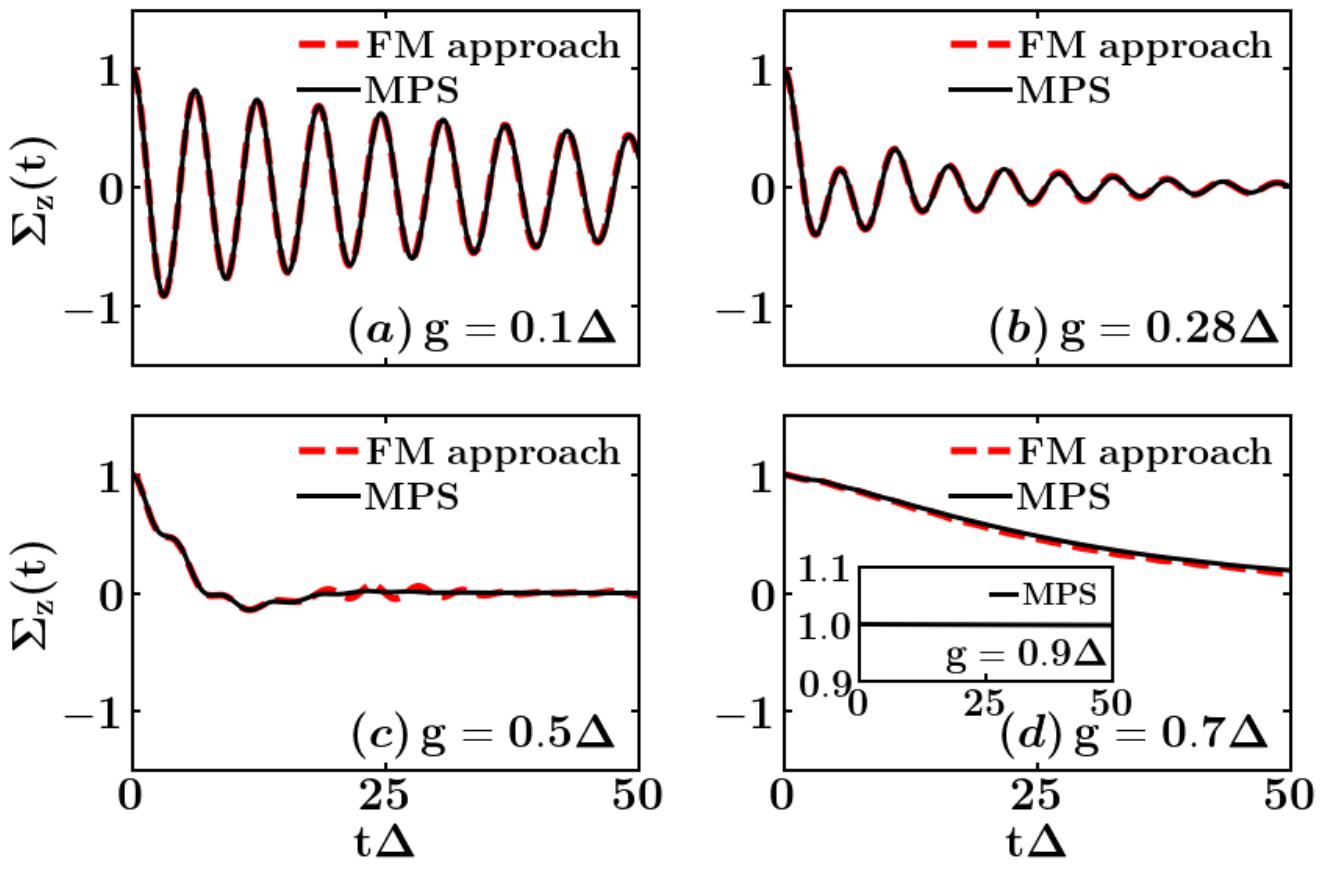}
  \caption{\label{fig:3} (color online)
    $\Sigma_z(t)$ at different values of $g/\Delta$: comparison between Feynman-Mori (FM) approach ($\beta \Delta=5000$) and MPS method ($T=0$). In the inset of panel (d) MPS simulation at $g \simeq g_c$, where there is no relaxation.}
\end{figure}

Furthermore $\Sigma_{z}(z)$, the Laplace-transformed relaxation function, is strictly related to the magnetic susceptibility $\chi(z)=-i \int_0^{\infty} e^{i z t} \langle [\sigma_z(t),\sigma_z(0)] \rangle dt$, where $z$ lies in the complex upper half plane, i.e. $z=\omega+i\epsilon$, with $\epsilon > 0$. Indeed $\Sigma_{z}(z)=i \frac{(\chi(z)-\chi(z=0))}{M^2\beta z}$, that is the analogue of the relation between the optical conductivity and the current-current correlation function in solids \cite{boltz}. By using the eigenbasis of the interacting system Hamiltonian and the commutation relation $[\sigma_z,H]=-i \Delta \sigma_y$, it is straightforward to deduce the following two very interesting properties:
\begin{eqnarray}
  M^2 \beta=-\frac{2}{\pi} \int_{0}^{\infty} \frac{\Im(\chi(\omega))}{\omega} d\omega,
\label{m2chi}
\end{eqnarray}
and
\begin{eqnarray}
  \Sigma_{z}(z)=\frac{i}{z}+\frac{(\sigma_y,\sigma_y)}{(\sigma_z,\sigma_z)} \Delta^2 \Sigma_{y}(z).
  \label{sigmayz}
\end{eqnarray}
Equation (\ref{m2chi}) shows that the behavior of the magnetic susceptibility at low frequencies is directly related to the order parameter of QPT. Note that
$M^2 \beta$, when $\beta \rightarrow \infty$, tends to a finite constant depending on $g$ for $g < g_c$, whereas, at $g \ge g_c$, diverges. On the other hand, equation (\ref{sigmayz}), which establishes a connection between $\Sigma_{z}(z)$ and $\Sigma_{y}(z)$, i.e. between the two relaxation functions along $z$ and $y$ axes, allows to define an effective gap: $\Delta_{eff}^2=\frac{(\sigma_y,\sigma_y)}{(\sigma_z,\sigma_z)} \Delta^2$. In particular it restores the bare qubit gap $\Delta$ at $g=0=\alpha_{cav}$. In Fig.~\ref{fig:2}d we plot the effective gap, in units of $\Delta$, as function of $g/\Delta$ at two different temperatures. We emphasize that this important physical quantity provides a precise indication of the onset of QPT, being related to $M^2$. Indeed $(\sigma_z,\sigma_z)=M^2$ and $(\sigma_y,\sigma_y)=\frac{2 \langle \sigma_x \rangle}{\beta \Delta}$. It is worth noticing that simple, but not accurate, variational approaches, based on polaronic unitary transformations \cite{zueco}, provide a discontinuity in the quantity $\langle H_{Q} \rangle$ that is generally associated to the onset of QPT. We highlight that this jump is an artifact of this kind of approximate methods, indeed it is present neither in WLMC technique nor the variational approach \`a la Feynman as previously discussed. It confirms that $\Delta_{eff}$, and then $M^2$, and not $\langle H_{Q} \rangle$ represents the right order parameter of QPT.

\begin{figure}[H]
  \includegraphics[width=1.01\columnwidth]{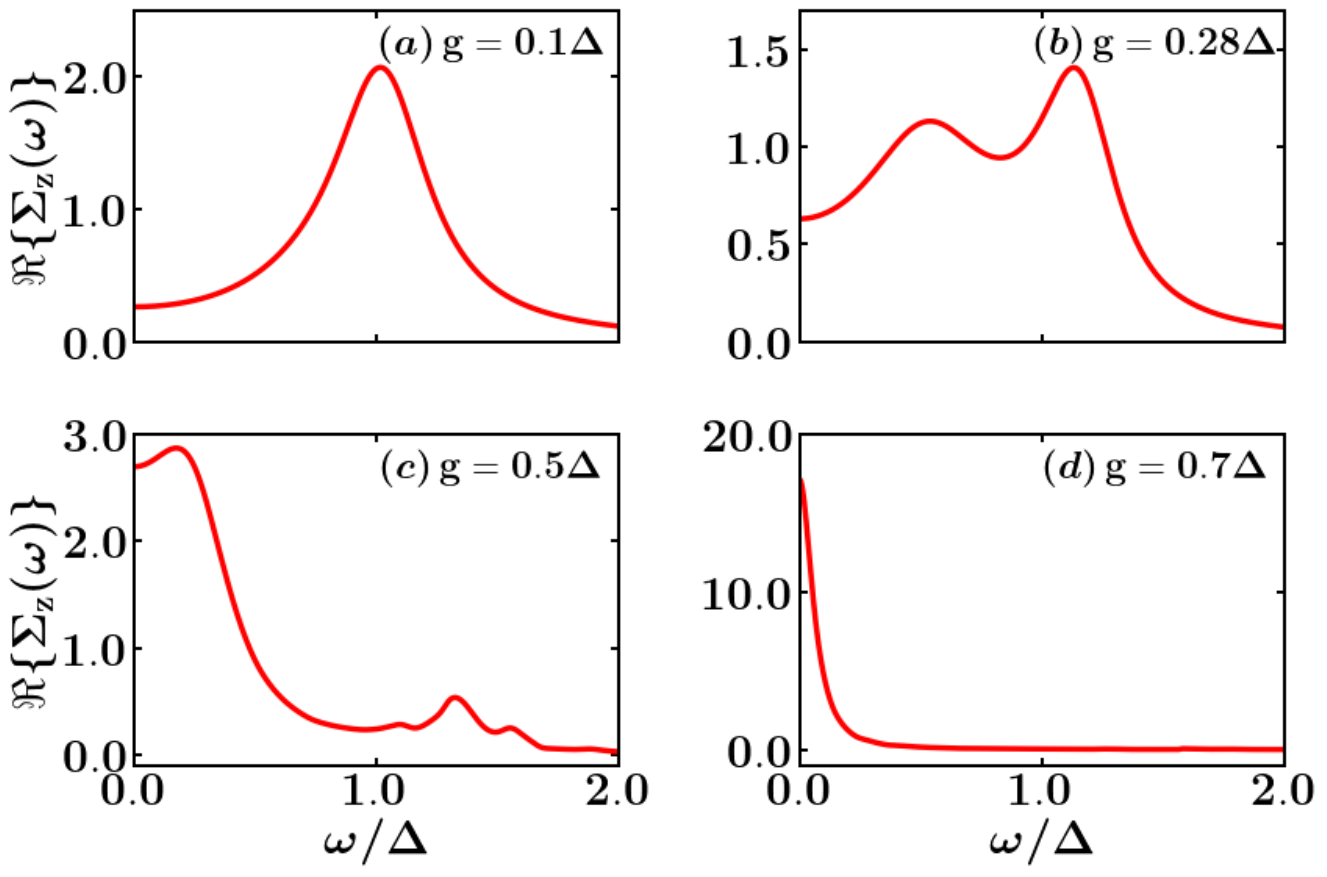}
  \caption{\label{fig:4} (color online)
    $\Sigma_z(\omega)$ at different values of $g/\Delta$ within Feynman-Mori approach ($\beta \Delta=5000$).}
\end{figure}

In a previous paper \cite{giulio2,boltz} we have proved that $\Sigma_{y}(z)$ can be exactly expressed in terms of a weighted sum contributions associated to the eigenstates of the interacting system, each characterized by its own frequency-dependent relaxation time:
\begin{eqnarray}
  \Sigma_{y}(z)=\sum_n P_{n} \frac{i}{z + i M_{n}(z)},
  \label{sigman}
\end{eqnarray}
with $\sum_n P_{n}=1$. We emphasize that so far there is no approximation. Here, we combine, for the calculation of $\Sigma_{y}(z)$, the short-time approximation, typical of the memory function formalism \cite{mori1}, and the approach \`a la Feynman, by replacing, in $M_{n}(z)$, the exact eigenstates of $H$ with that ones of $H_M$, whose parameters are variationally determined. Indeed, since the commutator between $\sigma_y$ and $H$ involves a contribution proportional to the qubit-boson coupling, the short-time approximation can be more easily implemented for the calculation of the relaxation function $\Sigma_{y}(z)$ \cite{giulio2}. Once $\Sigma_{y}(z)$ is known, Equation (\ref{sigmayz}) allows us to obtain $\Sigma_{z}(t)$. In Fig.~\ref{fig:3} we compare $\Sigma_{z}(t)$ with that obtained through MPS approach \cite{bnote}, where a standard matrix product operator representation of the time evolution operator $U(t+dt, t)=\exp(-i H d t)$ \cite{dmrg2} is implemented using the ITensor library \cite{fishman}. This method allows us to simulate the non-equilibrium dynamics of long-ranged model Hamiltonians starting from a generic initial state. In our case the initial state is the ground state of $H$ in the presence of a small magnetic field along $z$ axis, as previously discussed.

The plots in Fig.~\ref{fig:3} show that, at weak coupling, the dynamics is characterized by Rabi oscillations, whose amplitude and frequency reduce by increasing the strength of the coupling $g/\Delta$. By furtherly increasing $g/\Delta$, the relaxation becomes exponential: this is the analogue of the Toulouse point in the spin-boson model. Then the relaxation time gets longer and longer, and, at $g \ge g_c$, the system does not relax, i.e. $\Sigma_{z}(t)=1$ independently on time $t$, signalling the occurrence of QPT. The behavior of the relaxation function in the frequency domain sheds further light on the relaxation processes. In the weak coupling regime, the real part of $\Sigma_{z}(\omega)$ exhibits only a peak centered, essentially, at the bare qubit gap $\Delta$. At $g=0.28 \Delta$, the effective gap turns out to be equal to the resonator frequency: the spectrum presents avoided crossings, giving rise to the so called vacuum Rabi splitting \cite{fragner}. By increasing $g/\Delta$, there is a transfer of spectral weight towards lower frequencies. In particular, when $\Sigma_{z}(t)$ exhibits an exponential behavior, $\Re(\Sigma_{z}(\omega))$ is characterized by a peak centered at zero frequency. The width of this structure becomes narrower and narrower, and, at $g=g_c$, $\Re(\Sigma_{z}(\omega))$ exhibits a delta function centered at zero frequency: it signals the onset of QPT.

Starting from an inductive coupling between a flux qubit and an LC oscillator via Josephson junctions as in \cite{nature}, for the experimental observation we propose to introduce a dissipative element in the LC circuit. Following Devoret \cite{devo,lib} we replace it with a continuum of harmonic oscillators as in the Caldeira-Legget model \cite{newnote}. By using the values measured in Ref.~\cite{nature}, the resistance turns out to be $R \simeq \frac{0.24}{\alpha_{cav}}$ $k\Omega$, so that $\alpha_{cav} \simeq 0.2$ corresponds to $R \simeq 1.2$ $k\Omega$. Then, for moderate values of the dissipation, $R$ is of the order of $k\Omega$ and QPT occurs for values of $g_c/\omega_0 \simeq 1$, i.e. $g_c$ lies in the deep strong coupling regime that can be experimentally reached \cite{nature}.

{\it Conclusions.} We proved that the open quantum Rabi model exhibits a QPT by varying the strength of the coupling between the qubit and the resonator, even in the presence of extremely low dissipation magnitude. We characterized QPT unveiling its signatures both at and out of thermodynamic equilibrium by using typical linear response measurements. 

\bibliography{main_text}{}

\begin{thebibliography}{70}
\bibitem [{ano()}]{anote}%
  \BibitemOpen
  \href@noop {} {}\bibinfo {note} {In general, the strong coupling regime is
  obtained when $g$ exceeds the rate at which excitations decay into the
  environment, $g$ being the strength of the coupling qubit-resonator. If $g$
  is small with respect to the resonator frequency $\omega_0$ and the qubit gap
  $\Delta$, and the frequencies fulfill the condition $|\Delta-\omega_0| \ll
  |\omega_0 + \Delta|$, the interaction is faithfully described with the
  Jaynes-Cummings model, where coupling terms that simultaneously excite or
  de-excite the system and the field are neglected. When $g/\Delta$ and
  $g/\omega_0$ are about $0.1$, the Jaynes-Cummings model fails in rightly
  describing the Rabi model: this is the so called strong coupling regime.
  Finally if $g/\Delta$ and $g/\omega_0$ are of order of $1$ or greater than
  $1$, the deep strong coupling regime is reached.}\BibitemShut {Stop}%
\end{thebibliography}




%
%

\pagebreak
\widetext
\begin{center}
\textbf{\large  Supplementary Information for: Signatures of Dissipation Driven Quantum Phase Transition in Rabi Model}
\end{center}

\setcounter{equation}{0}
\setcounter{figure}{0}
\setcounter{table}{0}
\setcounter{page}{1}
\makeatletter

\section {The model.}
\label{sec: model}

The Hamiltonian can be written as sum of two contributions:
\begin{equation}\label{eq:definitionH}
	H=H_{Q-O} + H_{I},
\end{equation}
where: 1) $H_{Q-O}=-\frac{\Delta}{2}\sigma_{x}+\omega_0 a^\dagger a+ g \sigma_z (a+a^\dagger)$ describes the qubit-oscillator system, $\Delta$ being the tunneling matrix element, $a$ ($a^\dagger$) standing for the annihilation (creation) operator for the bosonic field with frequency $\omega_0$, and $g$ representing the strength of the coupling; 2) $H_{I}=\sum_{i=1}^N \left[ \frac{p_i^2}{2 M_i} + \frac{k_i}{2} (x-x_i)^2 \right]$ describes the environmental degrees of freedom and their interaction with the resonator. The bath is represented as a collection of harmonic oscillators with frequencies $\omega_i^2=\frac{k_i}{M_i}$, and coordinates and momenta given by $x_i$ and $p_i$, respectively; furthermore $x$ denotes the position operator of the resonator with mass $m$: $x=\sqrt{\frac{1}{2m\omega_0}}(a+a^\dagger)$. Units are such that $\hbar=k_B=1$. We emphasize that, in Eq.(\ref{eq:definitionH}), $\sigma_x$ and $\sigma_z$ are Pauli matrices with eigenvalues $1$ and $-1$. The dissipative environment is modeled as a strictly Ohmic bath with spectral density: $J(\omega)=\sum_{i=1}^N \frac{k_i \omega_i}{2m\omega_0}\delta(\omega-\omega_i)=\alpha_{cav}\omega \Theta(\omega_c-\omega)$. Here the adimensional parameter $\alpha_{cav}$ measures the strength of the coupling and $\omega_c$ is a cutoff frequency that is the largest energy scale, i.e. $\omega_c \gg \omega_0$ and $\omega_c \gg \Delta$. The first step is represented by the diagonalization of the Hamiltonian describing the cavity and its environment: $\omega_0 a^\dagger a + H_{I}$ \citeSupp{zuecosupp}. In particular, the eigenvalues, $\tilde{\omega}_i, i=1,..,N+1$, of the positive definite quadratic form representing the potential energy of the classical Lagrangian are given by the poles, on the real axis, of the bosonic field Green function:
\begin{equation}\label{eq:green}
	G(z)=\frac{2 \omega_0}{z^2-\omega_0^2 -A(z)},
\end{equation}
where $A(z)=\sum_{i=1}^N \frac{M_i}{m} \frac{\omega_i^2 z^2}{z^2-\omega_i^2}$ denotes the self-energy that can be exactly expressed in terms of the bath spectral density:
\begin{equation}\label{eq:selfenergy}
	A(z)=2 \omega_0 z^2 \int_{0}^{\infty} d\omega \frac{J(\omega)}{\omega \left ( z^2 -\omega^2 \right )}. 
\end{equation}

The diagonalization of the Hamiltonian describing the cavity and its environment leads to $N+1$ independent harmonic oscillators with frequencies $\tilde{\omega}_i$, i.e. an effective bath with Hamiltonian $H_{B}=\sum_{i=1}^{N+1} \tilde{\omega}_i c_i^\dagger c_i$. It allows us to map the original Hamiltonian to an effective model, where a single two level system, with gap $\Delta$, interacts, through $\sigma_z$ operator, with these renormalized bosonic modes with strength $\lambda_i$: $H_I^{\prime}=\sum_{i=1}^{N+1} \lambda_i \sigma_z (c_i+c_i^\dagger)$. It is straightforward to show that the effective spectral density function \citeSupp{zuecosupp}, $J_{eff}(\omega)=\sum_{i=1}^{N+1} \lambda_i^2 \delta(\omega-\tilde{\omega}_i)$, is strictly related to the imaginary part of the bosonic Green function $G(z)$:
\begin{equation}
	J_{eff}(\omega)=-g^2 \frac{\Im(G(\omega +i \epsilon))}{\pi},  
\end{equation}
with $\epsilon \rightarrow 0^{+}$. If the dissipative environment is modeled as a strictly Ohmic bath, $J(\omega)=\alpha_{cav} \omega \Theta(\omega_c-\omega)$, Eq.(\ref{eq:selfenergy}) provides:
\begin{equation}
	J_{eff}(\omega)=\frac{2 g^2 \omega_0^2\alpha_{cav}\omega}{(\omega^2-\omega_0^2-h(\omega))^2+(\pi \alpha_{cav} \omega_0 \omega)^2} \Theta(\omega_c-\omega),
	\label{spectralfunctionsupp}
\end{equation}
where $h(\omega)=\alpha_{cav} \omega_0 \omega \log \left [ \frac{\omega_c +\omega}{\omega_c - \omega}\right ]$.

$J_{eff}(\omega)$ features a Lorentzian peak at the oscillator frequency $\omega_0$ with width $\pi \alpha_{cav} \omega_0$, and, at low frequencies, $\omega \ll \omega_0$, exhibits an Ohmic behavior, $J_{eff}(\omega) \simeq \frac{\alpha_{eff}}{2} \omega$, with $\alpha_{eff}=4 g^2 \alpha_{cav} / \omega_0^2$. Furthermore $J_{eff}(\omega)$ practically vanishes at $\omega \simeq \omega_c$, so that the results are independent of the value of $\omega_c$, provided that $\omega_c \gg \omega_0$ and $\omega_c \gg \Delta$. In the following we fix $\Delta=1$, $\omega_0=0.75$ and $\omega_c=10$, as in the main text. \\

\newpage

\section  {The approaches.}

\subsection {The worldline Monte Carlo (WLMC) method.}

In the previous section, we proved, through an exact diagonalization, that it is possible to map the original Hamiltonian to an effective model, where a single two level system, with gap $\Delta$, interacts, through $\sigma_z$ operator, with $N+1$ renormalized bosonic modes, describing both the cavity and its environment:
\begin{equation}
	H_{eff}=-\frac{\Delta}{2}\sigma_{x}+\sum_{i=1}^{N+1} \tilde{\omega}_i c_i^\dagger c_i + \sum_{i=1}^{N+1} \lambda_i \sigma_z (c_i+c_i^\dagger),
\end{equation}
i.e. the original Hamiltonian is equivalent to an effective spin-boson Hamiltonian describing a two-level system interacting with a structured bath, whose effective spectral density is given by $J_{eff}(\omega)$.

WLMC is a path integral technique based on a Monte Carlo algorithm. The first step is the exact elimination of the bath degrees of freedom, that leads to an effective euclidean action \citeSupp{bulla1,weissup}:
\begin{equation}\label{eq:eqSsupp}
	S=\frac{1}{2}\int\limits_0^{\beta}\!d\tau\int\limits_0^{\beta}\!d\tau^{\prime} \sigma_z(\tau) K_{eff} (\tau-\tau^{\prime})\sigma_z(\tau^{\prime}),
\end{equation}
where the kernel is expressed in terms of the spectral density $J_{eff}(\omega)$ and the bath propagator:
\begin{equation}
	K_{eff}(\tau)=\int\limits_0^{\infty}\!d\omega J_{eff}(\omega) \frac{ \cosh \left[ \omega \left( \frac{\beta}{2}-\tau \right)\right]}{\sinh \left( \frac{\beta \omega}{2} \right)}.
\end{equation}

We emphasize that this step is exact, i.e. it contains not any approximation. Indeed, the exact elimination of the effective bath degrees of freedom is independent of the specific distribution of the bosonic frequencies and the particular form of the couplings $\lambda_i$. It is based only on the linearity of the coupling between the spin and its new effective environment describing both the cavity and original bosonic bath \citeSupp{feynman_new}.

The problem is then equivalent to a one-dimensional classical system of spin variables distributed on a chain with length $\beta$, and ferromagnetically interacting with each other ($\tau$ and $\tau^{\prime}$ label the spins on the chain). The next step is the calculation of the functional integral: it is carried out by weighting, with Poissonian measure, all the possible piecewise constant functions, i.e. the world-lines $\sigma_z(\tau)$, with values $1$ and $-1$, periodic of period $\beta=1/T$, where $T$ is the system temperature. An efficient sampling of the path integral can be performed adopting a cluster algorithm \citeSupp{riegersupp,bulla1} based on the  Swendsen \& Wang  approach \citeSupp{swangsupp}.

Starting from a world-line $\sigma_z(\tau)$, one introduces a number (not necessarily even) of potential flips, extracted from a Poissonian distribution with average $\beta\Delta/2$. Given the segments individuated by the real and potential flips, one connects two segments, with extremes $u_1$, $u_2$ and $u_3$, $u_4$, having the same value of the spin with the probability:
\begin{equation}
	p([u_1,u_2],[u_3,u_4])=1-\exp\left[-2\int\limits_{u_1}^{u_2}\!d\tau\int\limits_{u_3}^{u_4}\!d\tau^\prime\, K_{eff}(|\tau-\tau^\prime|)\right].
\end{equation}

Then one flips the connected clusters with probability $1/2$, and finally removes the flips that do not separate segments with a different value of the spin. We emphasize that this approach is exact from a numerical point of view, and it is equivalent to the sum of all the Feynman diagrams. \\

\subsection {The variational approach \`a la Feynman.}

This method is based on the variational principle and it is strictly related to the approach introduced by Feynman within the Fr\"ohlich model \citeSupp{Feynmansupp}. Here we resume the main steps. In general, in a many-body problem, one uses the Feynman-Dyson perturbation theory \citeSupp{fetter, Mahan} starting from the Hamiltonian without interactions, so the more the strength of the coupling increases, the more the number of Feynman diagrams to be considered becomes bigger. To overcome this difficulty, in the charge polaron problem, Feynman introduced, as starting point for the perturbation theory, a so smart variational action that the expansion to the first order is already enough to obtain an excellent description of the physics for arbitrary coupling strength. Here we follow this idea. We adopt the ordered operator calculus \citeSupp{Feynman1}, i.e. the operator equivalent of the Feynman path integral. The first step is the calculation of the partition function \citeSupp{fetter, Mahan} $Z=Tr \left[ e^{-\beta H}\right] =Z_{0} U(\beta)$, where $U(\beta)=\langle T_{\tau} e^{-\int_{0}^{\beta} d\tau^{\prime}H_{I}^{\prime,(0)}(\tau^{\prime})}\rangle_{0}$. Here $Z_{0}$ is the free partition function (related to $H_{0}=H_{Q}+H_{B}$), with $H_{Q}=-\frac{\Delta}{2}\sigma_{x}$ describing the qubit Hamiltonian, $T_{\tau}$ is the time ordering operator, $H_{I}^{\prime,(0)}$ represents the Hamiltonian $H_I^{\prime}$ in the interaction representation, and $\langle ... \rangle_{0}$ denotes the ensemble average with respect to $H_{0}$.  By choosing, for the trace, the basis of $H_{0}$ (it is factorized), it is possible to exactly eliminate the bath degrees of freedom by using the Bloch-DeDominicis theorem \citeSupp{fetter, Mahan}. The partition function becomes:
\begin{equation}\label{eq:zeta}
	Z=Z_{Q}Z_{B}\langle T_{\tau} e^{\phi}\rangle_{Q},
\end{equation}
where:
\begin{equation}\label{eq:phi}
	\phi=\frac{1}{2}\int_0^{\beta}d\tau \int_0^{\beta} d\tau^{\prime} \sigma_z^{(0)}(\tau) K_{eff}(\tau-\tau^{\prime}) \sigma_z^{(0)}(\tau^{\prime}),
\end{equation}
Note that, in Eq.(\ref{eq:zeta}), the ensemble average is with respect to $H_{Q}$ and that, differently from the path-integral representation, i.e. Eq.(\ref{eq:eqSsupp}), in Eq.(\ref{eq:phi}) spin operators and not their eigenvalues appear. So far no approximation has been used. In order to go ahead with the calculation, one can expand the exponential and use the standard perturbation theory in the many body problem. To avoid to evaluate a huge number of diagrams when the coupling with the bath increases, we follow the Feynman's idea and introduce a trial Hamiltonian ($H_{tr}$), where we replace the effective bath, characterized by a continuous distribution of harmonic oscillators, with a discrete collection of $M$ bosonic fictitious modes:
\begin{equation}\label{eq:definitionHtrial}
	H_{tr}=H_{Q} + \sum_{i=1}^{M}\tilde{\Omega}_{i} b_i^{\dagger}b_i+ \sigma_z \sum_{i=1}^M \tilde{\lambda}_i\left(b_i^\dagger+b_i\right),
\end{equation}
where the parameters $\tilde{\Omega}_{i}$ and $\tilde{\lambda}_{i}$ have to be variationally determined as specified in the following. Also in this case, due to the linearity of the coupling term, we can exactly eliminate the bosonic degrees of freedom, getting:
\begin{equation}\label{eq:zetatrial}
	Z_{tr}=Z_{Q} Z_{B_{tr}} \langle T_{\tau} e^{\phi_{tr}}\rangle_{Q},
\end{equation}
where
\begin{equation}\label{eq:phitrial}
	\phi_{tr}=\frac{1}{2}\int_0^{\beta}d\tau \int_0^{\beta} d\tau^{\prime} \sigma_z^{(0)}(\tau)K_{tr}(\tau-\tau^{\prime}) \sigma_z^{(0)}(\tau^{\prime}),
\end{equation}
and $K_{tr}(\tau)$ contains the propagator of these trial modes and the coupling strengths: $K_{tr}(\tau)=\sum_{i=1}^M \tilde{\lambda}_i^2 \frac{ \cosh \left [ \tilde{\Omega}_i \left( \frac{\beta}{2}-\tau \right)\right]}{\sinh \left( \frac{\beta \tilde{\Omega}_i}{2} \right)}$. Now it is straightforward to prove that the second derivative of the function $f(x)=-T \log \langle T_{\tau}e^{\phi_{tr}+x \left(\phi-\phi_{tr} \right)}\rangle_{Q}$ is negative for any value of $x$ in the range $[0,1]$ \citeSupp{bogoliubov}. This property gives rise to the following inequality: $f(x=1)-f(x=0)\le f^{\prime}(x=0)$, i.e. an upper bound for the free energy $F=-T \log Z$:
\begin{equation}\label{eq:freeenergy}
	F-F_B \le F_{tr}-F_{B_{tr}}-T \frac{ \langle T_{\tau} e^{\phi_{tr}} \left(\phi-\phi_{tr}  \right)\rangle_{Q}}{ \langle T_{\tau} e^{\phi_{tr}}\rangle_{Q}}.
\end{equation}
This is exactly the same inequality found by Feynman within the Fr\"ohlich model (Feynman-Jensen inequality)\citeSupp{Feynmansupp,Feynman2supp}: it is a generalization of the well known Bogoliubov inequality \citeSupp{Feynman2supp}. We emphasize that, in this variational formulation, only the free energy of the model Hamiltonian and the first correction enter. The knowledge of the eigenvalues and eigenvectors of $H_{tr}$, through numerical diagonalization, allows us to calculate the right side of Eq.(\ref{eq:freeenergy}). Indeed $\frac{ \langle T_{\tau} e^{\phi_{tr}} \left(\phi-\phi_{tr} \right)\rangle_{Q}}{ \langle T_{\tau} e^{\phi_{tr}}\rangle_{Q}}=\langle \phi-\phi_{tr} \rangle_{H_{tr}}$ \citeSupp{bogoliubov}. The variational procedure provides the values of the parameters $\tilde{\Omega}_{i}$ and $\tilde{\lambda}_{i}$. 

Finally, starting by the partial derivative of the free energy with respect to $\Delta$, $\omega_0$, and $g$, and replacing, at the end of the calculation, $\phi$ with $\phi_{tr}$, it is possible to obtain the average values of $\sigma_x$, $a^\dagger a$ (the average number of phonons of the resonator), and $g \sigma_z (a+a^\dagger)$ (the average value of the spin-resonator coupling term) respectively. In the following and in the main text we prove the effectiveness of this variational approach: $M=3$, i.e. three fictitious modes, are enough to obtain a successfull agreement with the WLMC technique up to $\beta=10000$. \\

\subsection {Mori approach.}

This approach allows us to study the relaxation towards the thermodynamic equilibrium. In general, Mori formalism permits to reformulate, in an exact way, the Heisenberg equation of motion of any dynamical variable in terms of a generalized Langevin equation \citeSupp{mori1supp}. First of all, we introduce a Hilbert space of operators (whose invariant parts are set to be zero) where the inner product is defined by $(A,B)= \frac {1}{\beta} \int_{0}^{\beta} \left\langle e^{sH} A^{\dagger} e^{-sH} B \right\rangle ds$. Any dynamical variable $O$ obeys the equation:
\begin{eqnarray}
	\frac{dO}{dt}=- \int_{0}^{t} M_O(t-t{'}) O(t{'}) dt{'} + f(t),
	\label{moriO}
\end{eqnarray}
where the quantity $f(t)$ represents the random force, that is, at any time, orthogonal to $O$ and is related to the memory function $M_O$ by the fluctuation-dissipation formula. The solution of this equation can be expressed as $O(t)=\Sigma_O(t) O+\tilde{O}(t)$, i.e. $\Sigma_O(t)=(O(t),O)/(O,O)$ describes the time evolution of the projection of $O(t)$ on the axis parallel to $O$ and represents the relaxation of the $O$ operator, whereas $\tilde{O}(t)$ is always orthogonal to $O$. The Laplace-transformed relaxation function, $\Sigma_O(z)=\int_0^{\infty} e^{i z t} \Sigma_O(t) dt$, where $z$ lies in the complex upper half plane, i.e. $z=\omega+i\epsilon$, with $\epsilon > 0$, can be exactly expressed either as $\Sigma_{O}(z)=\frac{i}{z+iM_{O}(z)}$, i.e. \`a la Mori, or in terms of a weighted sum of contributions associated to the exact eigenstates of the interacting system, each characterized by its own frequency-dependent relaxation time \citeSupp{boltzsupp}:
\begin{eqnarray}
	\Sigma_{O}(z)=\sum_n P_{n,O} \frac{i}{z + i M_{n,O}(z)}.
	\label{sigman}
\end{eqnarray}

Here:
\begin{eqnarray}
	P_{n,O}=\frac{e^{-\beta E_n}}{Z_p} \frac{\tilde{\Pi}_{n,O}}{C},
	\label{Pin}
\end{eqnarray}
where $C=\sum_n \tilde{\Pi}_{n,O} \frac{e^{-\beta E_n}}{Z_p}$, $Z_p$ is the partition function, $E_n$ are the exact eigenvalues of the Hamiltonian, $\omega_{n,m}=\frac{E_n-E_m}{2}$, and:
\begin{eqnarray}
	\tilde{\Pi}_{n,O}=-\mathop{\sum_{m}}_{E_m \ne E_n} \frac{\tanh(\beta \omega_{n,m})}{\omega_{n,m}} \left | \left\langle n \right | O \left | m \right\rangle \right |^2 -\beta \mathop{\sum_{m}}_{E_m=E_n} \left | \left\langle n \right | O \left | m \right\rangle \right |^2.
	\label{Pin1}
\end{eqnarray}

In particular, the weights $P_{n,O}$ are such that $\sum_n  P_{n,O}=1$. Furthermore, the memory functions associated to the eigenstates of the Hamiltonian are given by:
\begin{eqnarray}
	M_{n,O}(z)=i \frac{r_{n,O}(z)}{\frac{r_{n,O}(z)}{z}-\tilde{\Pi}_{n,O}},  
	\label{memoryn}
\end{eqnarray} 
where
\begin{eqnarray}
	r_{n,O}(z)=\mathop{\sum_{m}}_{E_m \ne E_n}\frac{\tanh(\beta \omega_{n,m})}{2 \omega_{n,m}} \left | \left\langle n \right | \tilde{O} \left | m \right\rangle \right |^2
	\left (\frac{1}{z+E_n-E_m}+\frac{1}{z-E_n+E_m} \right),
	\label{rn}
\end{eqnarray}
with $\tilde{O}=\frac{[O,H]}{i}$.

On the other hand it is straightforward to prove that the following commutation relation holds:
\begin{eqnarray}
	\frac{[\sigma_z,H]}{i}=-\Delta \sigma_y,
	\label{csz}
\end{eqnarray}

Then, by using the definitions of the two relaxation functions $\Sigma_z(t)$ and $\Sigma_y(t)$, i.e. $O=\sigma_z$ and $O=\sigma_y$, with $\Sigma_z(t)=(S_z(t),S_z)/(S_z,S_z)$ and $\Sigma_y(t)=(S_y(t),S_y)/(S_y,S_y)$, the definition of the inner product between two observables, and Eq.(\ref{csz}), it is possible to prove that $\frac{d^2 \Sigma_z(t)}{dt^2}=-\frac{(S_y,S_y)}{(S_z,S_z)} \Delta^2 \Sigma_{y}(t)$. It implies the following relation between the Laplace-transformed relaxation functions: 
\begin{eqnarray}
	\Sigma_{z}(z)=\frac{i}{z}+\frac{(S_y,S_y)}{(S_z,S_z)} \Delta^2 \Sigma_{y}(z).
	\label{sigmayzsupp}
\end{eqnarray}
In particular Eq.(\ref{sigmayzsupp}) allows us to define the effective gap: $\Delta_{eff}^2=\frac{(S_y,S_y)}{(S_z,S_z)} \Delta^2$. 

We emphasize that so far there is no approximation. Here, we combine, for the calculation of $\Sigma_{y}(z)$, the short-time approximation, typical of the memory function formalism \citeSupp{mori1supp}, and the approach \`a la Feynman, by replacing the exact eigenstates of $H$ with that ones of $H_{tr}$, whose parameters are variationally determined. Indeed, we start from the consideration that commutator between $\sigma_y$ and $H$ involves a contribution proportional to the spin-boson coupling:
\begin{eqnarray}
	\frac{[\sigma_y,H]}{i}=T+ 2 \sigma_x \sum_{i=1}^{N+1} \lambda_i (c_i+c_i^\dagger),
	\label{csy}
\end{eqnarray}
$T$ being an operator involving only the spin variables, i.e. it does not depend on the operators describing the effective bath. Then the short-time approximation can be more easily implemented for the calculation of the relaxation function $\Sigma_{y}(z)$. The function $\Sigma_{z}(z)$ can be obtained through Eq.(\ref{sigmayzsupp}). 

The first step is to consider $O=\sigma_y$ and to rewrite $M_{n,y}=M_{n,y}^{(M)}+\left( M_{n,y}-M_{n,y}^{(M)}\right)$, where $M_{n,y}^{(M)}$ is the memory function corresponding to the model Hamiltonian. Next, we use the short-time approximation for calculating the quantity $(M_{n,y}-M_{n,y}^{(M)})$, i.e., in Eq.(\ref{memoryn}), the function $\frac{r_{n,y}}{z}$ is neglected with respect to $\tilde{\Pi}_{n,y}$. Finally, just as in the variational approach \`a la Feynman, we calculate perturbatively the effect of the spin-boson coupling, i.e. we treat at the lowest order the correction with respect to the model Hamiltonian. This approach provides $M_{n,y}(z)=\frac{i}{z} \Delta^2 v_{n}(z)-\frac{i}{\tilde{\Pi}_{n,y}} ( f_n(z)-g_n(z) ) $, with:
\begin{eqnarray}
	v_{n}(z)_=\frac{ \mathop{\sum_{m}}_{E_m \ne E_n} \tanh(\beta \omega_{n,m}) \omega_{n,m} \left | \left\langle n \right | \sigma_y \left | m \right\rangle \right |^2 \frac{1}{z^2-(E_m -E_n)^2} } {\mathop{\sum_{m}}_{E_m \ne E_n} \tanh(\beta \omega_{n,m}) \omega_{n,m} \left | \left\langle n \right | \sigma_z \left | m \right\rangle \right |^2 \frac{1}{z^2-(E_m -E_n)^2 }},
	\label{vn}
\end{eqnarray}
\begin{eqnarray}
	\begin{split}
		f_{n}(z)_=\sum_{m} \left | \left\langle n \right | \sigma_x \left | m \right\rangle \right |^2 \int_{0}^{\infty} d\omega J_{eff}(\omega) ( & \frac {\tanh(\frac{\beta}{2}\left[E_m-E_n-\omega \right])}{E_m-E_n-\omega} N_{\omega} \left[ \frac{1}{z+E_n-E_m+\omega}+ \frac{1}{z-E_n+E_m-\omega} \right] + \\
		& \frac {\tanh(\frac{\beta}{2}\left[E_m-E_n+\omega \right])}{E_m-E_n+\omega} (N_{\omega}+1) \left[ \frac{1}{z+E_n-E_m-\omega}+ \frac{1}{z-E_n+E_m+\omega} \right]),
		\nonumber
	\end{split}
\end{eqnarray}
and
\begin{eqnarray}
	\begin{split}
		g_{n}(z)_=\sum_{m} \left | \left\langle n \right | \sigma_x \left | m \right\rangle \right |^2  \sum_{i=1}^{M} \tilde{\lambda}_i^2
		( & \frac {\tanh(\frac{\beta}{2}\left[E_m-E_n-\tilde{\Omega}_i \right])}{E_m-E_n-\tilde{\Omega}_i} N_{\tilde{\Omega}_i} \left[ \frac{1}{z+E_n-E_m+\tilde{\Omega}_i}+ \frac{1}{z-E_n+E_m-\tilde{\Omega}_i} \right] + \\
		& \frac {\tanh(\frac{\beta}{2}\left[E_m-E_n+\tilde{\Omega}_i \right])}{E_m-E_n+\tilde{\Omega}_i} (N_{\tilde{\Omega}_i}+1) \left[ \frac{1}{z+E_n-E_m-\tilde{\Omega}_i}+ \frac{1}{z-E_n+E_m+\tilde{\Omega}_i} \right]).
		\nonumber
	\end{split}
\end{eqnarray}
Here $N_{\omega}=\frac{1}{e^{\beta \omega}-1}$ represents the average number of phonons of the mode with frequency $\omega$. In the main text we proved that this approach is able to recover the numerically exact data got through MPS simulations. 

Now we want to focus our attention on two exact expressions regarding the relation between the effective gap and squared magnetization and the relation between the relaxation function and the magnetic susceptibility. First of all, by using the definition of the inner product between two operators, Eq.(\ref{csz}), and the commutation relation between $\sigma_y$ and $\sigma_z$, i.e. $[\sigma_y,\sigma_z]=2 i \sigma_x$, it is straightforward to prove that:
\begin{eqnarray}
	(\sigma_y,\sigma_y)=\frac{2 \langle \sigma_x \rangle}{\beta \Delta}, 
	\nonumber
\end{eqnarray}
and then
\begin{eqnarray}
	\Delta_{eff}^2=\frac{(S_y,S_y)}{(S_z,S_z)} \Delta^2= \Delta \frac{2 \langle \sigma_x \rangle}{\beta M^2}.
	\nonumber
\end{eqnarray}
Indeed $(\sigma_z,\sigma_z)=M^2=\frac{1}{\beta}\int_{0}^{\beta} d\tau \langle \sigma_z(\tau) \sigma_z(0)\rangle$. On the other hand the quantity $\beta M^2$, when $\beta \rightarrow \infty$, represents the correlation length of the 1D mapped model describing spins ferromagnetically interacting with each other, so that it tends to a finite constant depending on $g$ for $g < g_c$, whereas, at $g \ge g_c$, diverges. Here $g_c$ represents the critical value of the coupling with the resonator: at $g_c$ a Beretzinski-Kosterlitz-Thouless (BKT) quantum phase transition occurs, as we have discussed in the main text. Then, at zero temperature, the effective gap $\Delta_{eff}$ shrinks as function of $g$ and vanishes at $g=g_c$.  

The other important relation involves the Laplace-transformed relaxation function $\Sigma_z(z)$ and the magnetic susceptibility $\chi(z)$, that represents a typical linear response measurement. Here 
\begin{eqnarray}
	\chi(z)=-i \int_0^{\infty} e^{i z t} \langle [S_z(t),S_z(0)] \rangle dt.
	\label{chiz}
\end{eqnarray}
Indeed by using the eigenbasis of the interacting system Hamiltonian, it is straightforward to show that:
\begin{eqnarray}
	\Sigma_z(z)=i \frac{(\chi(z)-\chi(z=0))}{M^2 \beta z}.
	\label{chisigma}
\end{eqnarray}
Eq.(\ref{chisigma}) is the analogue of the relation between the optical conductivity and the current-current correlation function in solids \citeSupp{boltzsupp}.

Finally, by taking into account that $\chi(z)=\mathop{\sum_{m}}_{E_m \ne E_n} \frac{\left | \left\langle n \right | \sigma_x \left | m \right\rangle \right |^2}{z+E_n-E_m} \frac{(e^{-\beta E_n}-e^{-\beta E_m})}{Z_p}$ it is straightforward to prove that:
\begin{eqnarray}
	M^2 \beta=-\frac{2}{\pi} \int_{0}^{\infty} \frac{\Im(\chi(\omega))}{\omega} d\omega.
	\label{m2chisupp}
\end{eqnarray}
This equation establishes the relation between the magnetic susceptibility and the order parameter of the BKT quantum phase transition. 

\subsection {Matrix Product State simulations: qubit relaxation.}

In the main text we have used time-dependent Matrix Product State (MPS) simulations to study the qubit relaxation solving the mapped Hamiltonian: $H=H_Q+H_B+H_I^{\prime}$. In particular, we adopted the star geometry to describe the long-range interactions in $H_I^{\prime}$ between the qubit and the bath modes.\\

Because of the long-range character of these interactions, we used the method developed in Ref.~\citeSupp{longrange} for the solution of the time-dependent Schr\"odinger equation. It has been implemented using the ITensor library \citeSupp{ITensor}, to which we refer to as $W^I$. It consists in a first order approximation of the unitary time-evolution operator in terms of a Matrix Product Operator (MPO). This method has an error per site which diverges with the system size $L$, while giving a time-step error of $O(dt^2)$.\\

In the star-geometry considered in this work we placed the qubit of frequency $\Delta$ on the first site and on the remaining sites the collection of $N+1=600$ bosonic modes of the bath with frequencies $\tilde{\omega}_i$, each with Hilbert space of dimension $N_{ph}$. The couplings between the qubit and each bosonic mode $\lambda_i$ are defined such that we can describe the bath in terms of the effective spectral density $J_{eff}(\omega)$.\\

\begin{figure}[H]
	\centering
	\includegraphics[scale=0.325]{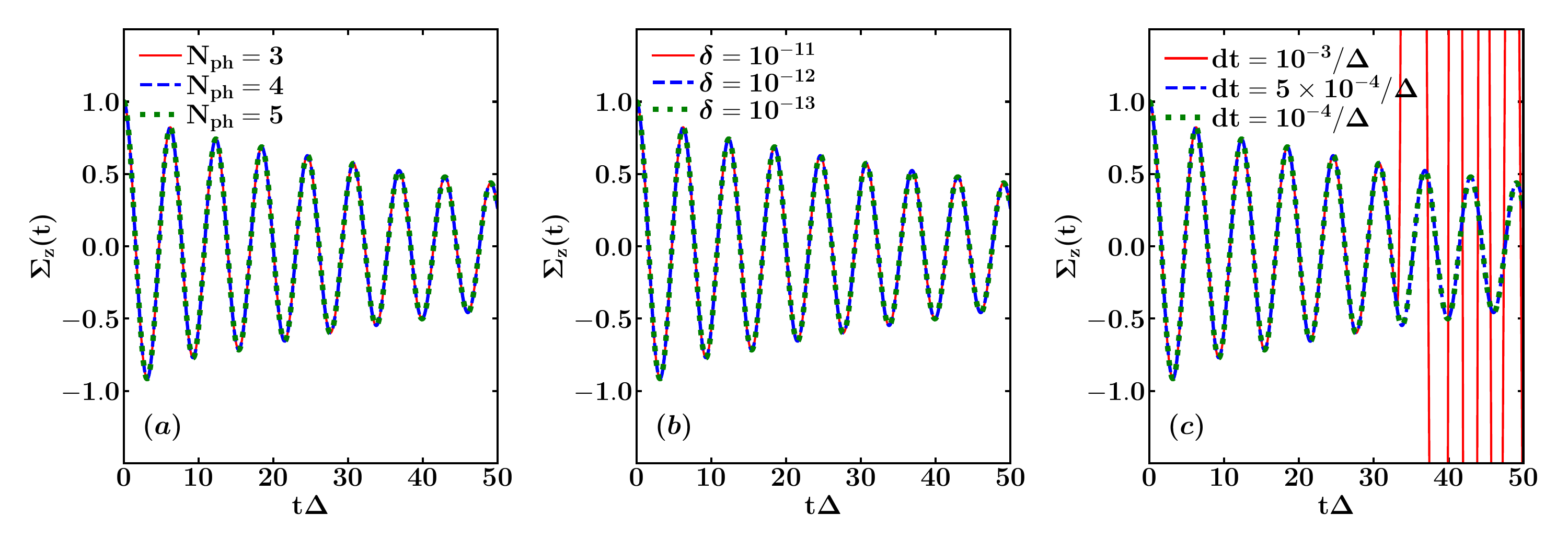}
	\caption{\label{convergence}Qubit relaxation as a function of time for the qubit with frequency $\Delta$ taken unitary, so that the cavity frequency is $\omega_0=0.75$, the coupling between them is $g=0.1$ and that between cavity and bath is $\alpha_{cav}=0.2$. In panel $(a)$ the time step is $dt=10^{-4}/\Delta$ and the cutoff is $\delta=10^{-13}$, while the number of phonons on each mode of the bath varies from $N_{ph}=3$ to $N_{ph}=5$; in panel $(b)$ the time step is $dt=10^{-4}/\Delta$ and the number of phonons on each mode of the bath is $N_{ph}=4$, while the cutoff $\delta$ varies from $10^{-13}$ to $10^{-11}$; in panel $(c)$ the cutoff is $\delta=10^{-13}$ and the number of phonons on each mode of the bath is $N_{ph}=4$, while the time step varies from $dt=10^{-3}/\Delta$ to $dt=10^{-4}/\Delta$.}
\end{figure}

As described in the main text, to simulate the relaxation dynamics of the system from the thermal equilibrium state at $T = 0$, we chose as initial state of the time evolution the ground state of the Hamiltonian $\mathcal{H}^{\prime} = \mathcal{H} + \epsilon\sigma_z$,where $\epsilon = 10^{-3}\Delta$ is a small magnetic field applied along $z$ axis. This ground state was computed by employing the DMRG algorithm, whose results were compared with those obtained through the Feynman approach and Worldline Monte Carlo numerical simulations. We converged our simulations of the relaxation of the qubit magnetization in the number of the Fock states of the bath modes, finding the best compromise between the smallest bond dimension and longest simulation times. We studied the behaviour of the qubit relaxation for different values of the qubit-oscillator coupling $g$ in the range $[0.1\Delta, 0.9\Delta]$, fixing $\alpha_{cav} = 0.2$ and $\omega_c = 10\Delta$. Our truncation error was kept below $\delta=10^{-13}$, requiring a maximum bond dimension of $D_{\rm {max}}=50$ and a time step $dt=5\times 10^{-4}/\Delta$ as shown in Fig.~\ref{convergence}. At the same time, this optimal maximum bond dimension allowed us to reach a final time for our simulations as big as $t_{\rm {final}} = 50/\Delta$. \\

\subsection {Matrix Product State simulations: cavity relaxation.}

\begin{figure}[H]
	\centering
	\includegraphics[scale=0.8]{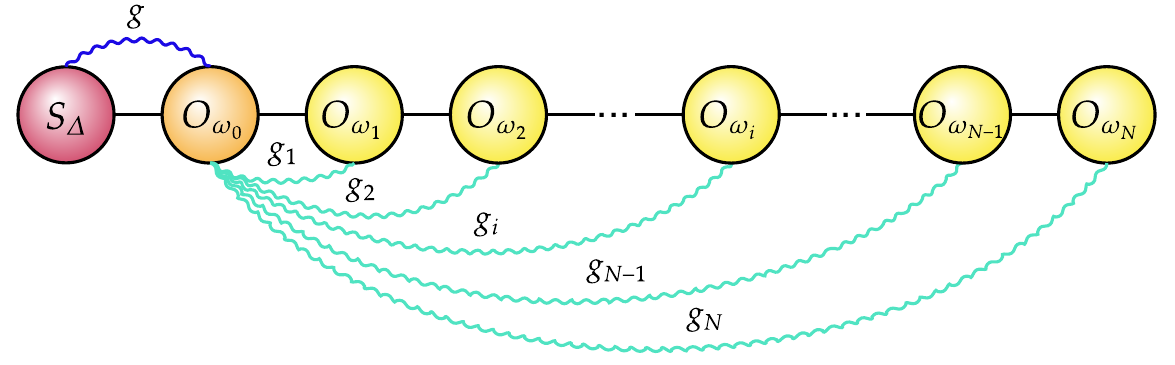}
	\caption{\label{fig:MPS} MPS chain of sites representing the Hamiltonian. The first site is occupied by the qubit, the second one by the resonator with Hilbert space of dimension $N_o$ and the sites from the third to the $(N+2)$th form the Ohmic bath of bosonic modes, each with Hilbert space of dimension $N_{ph}$. Qubit and resonator are coupled via $g$, while the resonator and the N bath modes are coupled by long range interactions with strength $g_i$.}
\end{figure}

In the following we will investigate the physical features of the relaxation function involving the resonator position operator. In this case we are forced to solve the original Hamiltonian, Eq.(\ref{eq:definitionH}). We adopt again the star geometry to describe the long-range interactions between the cavity and the bath modes. As depicted in Fig.~\ref{fig:MPS}, we place the qubit on the first site, on the second one the resonator, whose Hilbert space has dimension $N_o$, and on the remaining sites the collection of $N$ bosonic modes of the bath, each described by Hilbert space of dimension $N_{ph}$. The resonator and the bath experiment long-range interactions with couplings $g_i=k_i\omega_i/(2m\omega_0)$, with $i=1, ..., N$. We study the cavity relaxation starting the time evolution from the ground state of the Hamiltonian $\mathcal{H}^{\prime\prime} = \mathcal{H} + \gamma \frac{(a+a^{\dagger})}{\sqrt{2 m \omega_0}}$, where $\gamma$ is proportional to a small electric field acting on the resonator and $\frac{\gamma}{\sqrt{2 m \omega_0}} \simeq 10^{-3} \Delta$. Again, this ground state is computed by employing the DMRG algorithm.\\

In this case, the convergence of the numerical simulations becomes more complex. Hence, we decided to adopt the time-dependent variational principle (TDVP) \citeSupp{Haegeman2011,Haegeman2016,Paeckel2019}, where the time-dependent Schr\"odinger equation is projected to the tangent space of the MPS manifold of fixed bond dimension at the current time. In this work we employ the two-site TDVP (2TDVP in Ref.~\citeSupp{Paeckel2019}) using the second order integrator by sweeping left-right-left with half time step $dt/2$. This method shows a smaller time-step error $O(dt^3)$, and its accuracy is controlled only by the MPS bond dimension and the threshold to terminate the Krylov series. Here we stop the Krylov vectors recurrence when the total contribution of two consecutive vectors to the matrix exponential is less than $10^{-12}$. We emphasize that the TDVP method gives the correct solution by using a time step two or three orders of magnitude larger than that needed by $W^I$. Moreover, as expected, $W^I$ shows a much smaller accuracy than the TDVP method. These are the reasons why we adopt TDVP method for our simulations. An important observation about our numerical simulations is in order: we do not use more sophisticated approaches like local basis optimization \citeSupp{Zhang1998,Bursill1999,Friedman2000,Wong2008,Brockt2015,jansen2022finite}. We instead converge our simulations in the number of Fock states in the cavity ($N_o=60$) and bath modes ($N=500$), finding the best compromise between the smallest bond dimension and longest simulation times. Our truncation error is kept below $10^{-13}$ requiring a maximum bond dimension of $D_{\rm {max}}=50$. At the same time, this optimal maximum bond dimension allows us to reach a final time for our simulations as big as $t_{\rm {final}} \simeq 50/\Delta$. \\

We finally note that, in the star geometry, one could also adopt the TEBD method with swap gates. It was recently shown, however, in Ref.~\citeSupp{Liu2022}, that it usually requires larger bond dimensions compared to 2TDVP, despite giving smaller accumulated errors for long time evolutions.\\

\section  {BKT vs Mean Field phase transition.}

Here we address the physical features of the two phase transitions occurring in the dissipative quantum Rabi model. The first step is the exact elimination of the bosonic degrees of freedom: Eq.(\ref{eq:eqSsupp}) shows that the original problem turns to be equivalent to one-dimensional classical system of spins distributed on a chain with length $\beta$, and ferromagnetically interacting with each other. In particular, if $\omega_0$ is kept constant and the length tends to infinity, i.e. $\beta\rightarrow\infty$, it is straightforward to prove that the kernel has the following asymptotic behavior: $K_{eff}(\tau)=\frac{\alpha_{eff}}{2 \tau^2}$. It determines, by varying $g$, the occurrence of BKT quantum phase transition at a critical value $g_c$ of the coupling strength. In the main text we demonstrated that the order parameter, the squared magnetization $(\sigma_z,\sigma_z)=M^2=\frac{1}{\beta}\int_{0}^{\beta} d\tau \langle \sigma_z(\tau) \sigma_z(0)\rangle$, exhibits a jump at $g_c$. The question we address here is the following: what are the physical consequences on the resonator of the BKT quantum phase transition occurrence?

To this aim we introduce the Matsubara Green function of the resonator:
\begin{eqnarray}
	D(\tau)=-\langle T_{\tau} x(\tau)x(0) \rangle=-\frac{1}{2 m \omega_0} \langle T_{\tau} A(\tau)A(0) \rangle,
	\label{greenresonator}
\end{eqnarray}
where $A=a+a^{\dagger}$, $A(\tau)=e^{H \tau} A e^{-H \tau}$ and $0 <\tau <\beta$. On the other hand, by using that Green functions can be obtained through differentiation of the generating functional based on the partition function, it is straightforward to show that:
\begin{eqnarray}
	\langle T_{\tau} A(\tau_1)A(\tau_2) \rangle= \frac {\partial^2}{\partial s \partial t } \frac{ \langle T_{\tau} e^{\tilde{\phi}}\rangle_{Q}}{\langle T_{\tau} e^{\phi}\rangle_{Q}} |_{s=t=0},
	\label{greender}
\end{eqnarray}
where $\phi$ is given by Eq.(\ref{eq:phi}) and
\begin{equation}\label{eq:phitilde}
	\tilde{\phi}=\frac{1}{2}\int_0^{\beta}d\tau \int_0^{\beta} d\tau^{\prime} (g \sigma_z^{(0)}(\tau) +s \delta(\tau-\tau_1)+ t \delta(\tau-\tau_2)) \frac{K_{eff} (\tau-\tau^{\prime})}{g^2} (g \sigma_z^{(0)}(\tau^{\prime})+ s \delta(\tau^{\prime}-\tau_1)+ t \delta(\tau^{\prime}-\tau_2)).
\end{equation}
Eq.(\ref{greender}) provides:
\begin{eqnarray}
	D(\tau_1-\tau_2)=D^{(0)}(\tau_1-\tau_2)+\int_0^{\beta}d\tau \int_0^{\beta} d\tau^{\prime} D^{(0)}(\tau_1-\tau) g^2 (2 m \omega_0) \chi(\tau-\tau^{\prime}) D^{(0)}(\tau^{\prime} -\tau_2),
	\label{dtau1tau2}
\end{eqnarray}
where $\chi(\tau)$ is the Matsubara magnetic susceptibility and $D^{(0)}(\tau_1-\tau_2)$ is the boson Green function at $g=0$, i.e. in the absence of spin-resonator coupling:
\begin{eqnarray}
	D^{(0)}(\tau)=-\frac{1}{2 m \omega_0} \int\limits_0^{\infty}\ d\omega \frac{J_{eff}(\omega)}{g^2} \frac{ \cosh \left[ \omega \left( \frac{\beta}{2}-\tau \right)\right]}{\sinh \left( \frac{\beta \omega}{2} \right)}.
	\label{d0}
\end{eqnarray}
By considering the Fourier coefficients of both sides of Eq.(\ref{dtau1tau2}) one obtains:
\begin{eqnarray}
	D(i \omega_n)=D^{(0)}(i \omega_n)+2 m \omega_0 g^2 D^{(0)}(i \omega_n) \chi (i \omega_n) D^{(0)}(i \omega_n),
	\label{diomn}
\end{eqnarray}
$i \omega_n$ being Matsubara frequencies. In particular, at $i \omega_n=0$, Eq.(\ref{diomn}) becomes:
\begin{eqnarray}
	X^2=C+ 2 m \omega_0 g^2 \beta^2 C^2 M^2,
	\label{x2}
\end{eqnarray}
where $X^2=(x,x)=\frac{1}{2 m \omega_0} \frac{1}{\beta} \int_{0}^{\beta} d\tau \langle A(\tau) A(0)\rangle$ and $C=\frac{1}{\beta} \frac{1}{2 m \omega_0} \int\limits_0^{\infty}\ d\omega \frac{J_{eff}(\omega)}{g^2} \frac{2}{\omega}$. It is evident that the quantity $X^2$ for the resonator is the equivalent of $M^2$ for the two-level system. When BKT quantum phase transition sets in, $M^2$ and then $X^2$ exhibit a discontinuity. In other terms, for $g \ge g_c$ and $T=0$, both $\langle \sigma_z \rangle $ and $\langle x \rangle $ are not zero due to the spontaneous symmetry breaking.   

\begin{figure}[H]
	\centering
	\includegraphics[scale=0.4]{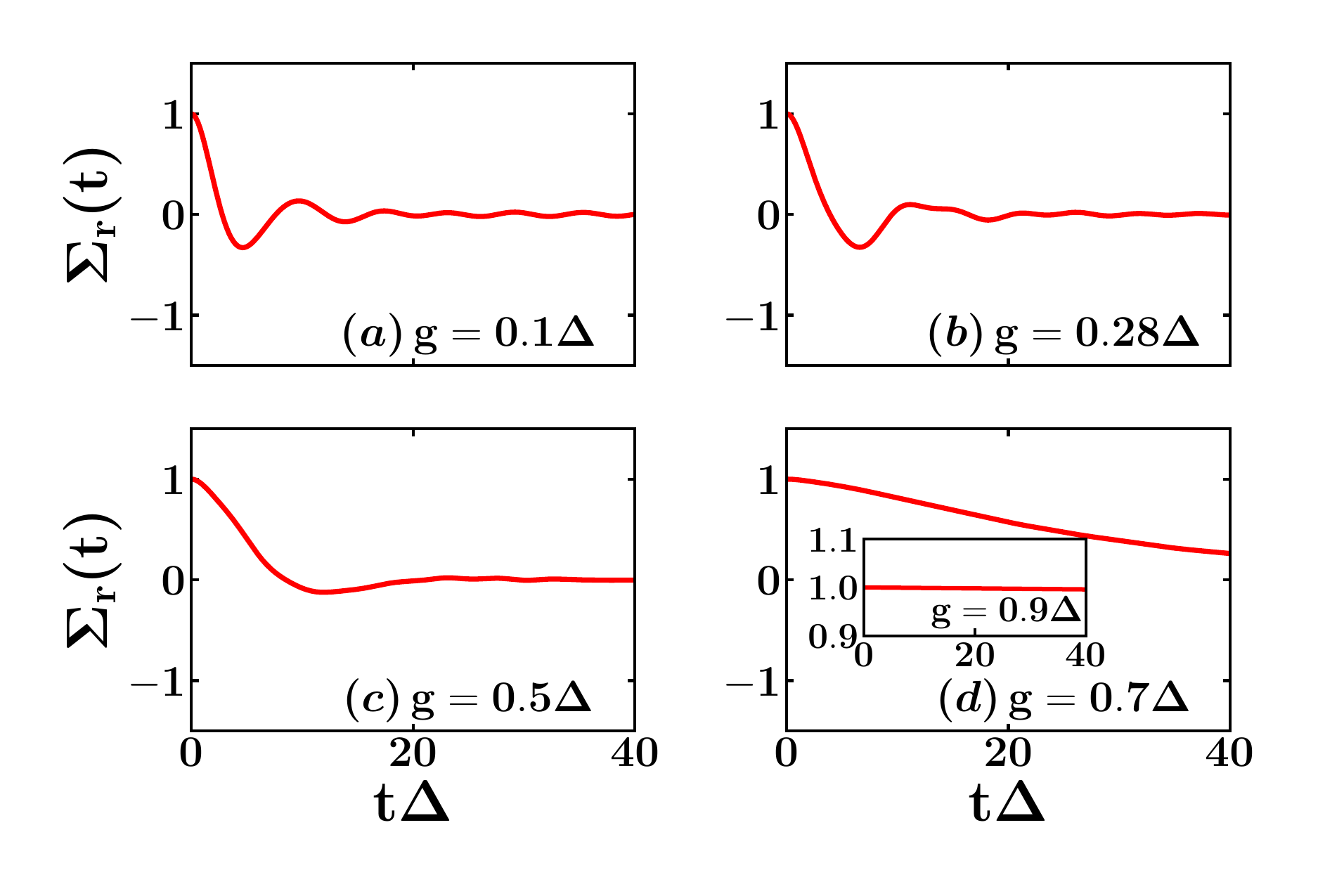}
	\caption{\label{fig:rmps}
		$\Sigma_r(t)$ at different values of $g/\Delta$: MPS method ($T=0$, $\omega_0=0.75$ and $\omega_c=10$). In the inset of panel (d) MPS simulations at $g \simeq g_c$, where there is no relaxation.}
\end{figure}

As in the main text, where we studied the relaxation of the $\sigma_z$ operator after applying a small magnetic field along $z$ axis, here we investigate the relaxation of the resonator position operator. Let's suppose that the system at $t=-\infty$ is at the thermal equilibrium. The response of the system to a perturbation, adiabatically applied from $t=-\infty$ and cut off at $t=0$, can be calculated within the Mori formalism and the linear response theory. In particular, here we adibatically apply a small electric field $E$ acting on the resonator. The most important physical quantity $\forall t \geq 0$ is the relaxation function $\Sigma_{r}(t)=\frac{(x(t),x(0))}{(x(0),x(0))}=\frac{\langle x(t) \rangle}{\langle x(0) \rangle}$ (calculated in the absence of $E$, being $t \geq 0$). In Fig.~\ref{fig:rmps} we plot $\Sigma_{r}(t)$: it is obtained through MPS simulations as previously described.  

The plots in Fig.~\ref{fig:rmps} show that, already at $g=0$, the Rabi oscillations of the resonator are damped due to the interaction with the environment. The amplitude and frequency of these oscillations furtherly reduce by increasing the strength of the coupling $g/\Delta$. Then the relaxation becomes exponential with the relaxation time getting longer and longer, and, at $g \ge g_c$, the system does not relax, i.e. $\Sigma_{r}(t)=1$ independently on time $t$, signalling the occurrence of quantum phase transition.

Now we focus our attention on the other phase transition, i.e. the superradiant phase transition, occurring both in the Rabi \citeSupp{ashhabsupp,plenio1supp} and dissipative Rabi model \citeSupp{plenio2supp}. We begin our discussion with the closed Rabi model:
\begin{eqnarray}
	H_{Q-O}=-\frac{\Delta}{2}\sigma_{x}+\omega_0 a^\dagger a+ g \sigma_z (a+a^\dagger).
	\label{closedrabi}
\end{eqnarray}
The resonator is represented by a harmonic oscillator with frequency $\omega_0$, mass $m$ and stiffness $k=m\omega_0^2$.

The exact elimination of the resonator degrees of freedom leads to an effective euclidean action:
\begin{equation}\label{sclosedrabi}
	\tilde{S}=\frac{1}{2}\int\limits_0^{\beta}\!d\tau\int\limits_0^{\beta}\!d\tau^{\prime\
	} \sigma_z(\tau)\tilde{K}(\tau-\tau^{\prime})\sigma_z(\tau^{\prime}),
\end{equation}
where the kernel is expressed in terms of the resonator propagator:
\begin{equation}
	\tilde{K}(\tau)=g^2 \frac{ \cosh \left[ \omega_0 \left( \frac{\beta}{2}-\tau \right)\right]}{\sinh \left( \frac{\beta \omega_0}{2} \right)}.
	\label{kclosedrabi}
\end{equation}

Also in this case, the problem is equivalent to a one-dimensional classical system of spin variables distributed on a chain with length $\beta$, and ferromagnetically interacting with each other, with an exponentially decaying coupling. However, when $\omega_0 \rightarrow 0$ and $\beta \rightarrow \infty$ with $\beta \omega_0 \rightarrow 0$, the spins experience a long-range interaction, whose strength is independent of the distance: $\tilde{K}(\tau) \rightarrow \frac{2 g^2}{\beta \omega_0}$. It is then clear that, under the assumptions $\omega_0 \rightarrow 0$ and $\beta \rightarrow \infty$, a ferromagnetic mean field phase transition occurs.

The same conclusion can be reached by using a variational approach that becomes exact in the adiabatic limit, i.e. $\omega_0 \rightarrow 0$, $m \rightarrow \infty$ keeping constant the stiffness $k=m \omega_0^2$. Indeed, in this limit, the resonator cannot follow the spin oscillations and the wave function of the system can be factorized into a product of normalized variational functions $\left |\varphi \right\rangle$ and $\left |f \right\rangle$, depending on the spin and bosonic coordinates respectively:
\begin{equation}
	\left | \psi \right\rangle =\left | \varphi \right\rangle \left |f \right\rangle.
	\label{eq:ad1}
\end{equation}

The expectation value of the Hamiltonian on the state $\left | \varphi \right\rangle $ provides an effective Hamiltonian for the resonator whose ground state is a coherent state:
\begin{equation}
	\left | f \right \rangle =e^{\left( \frac{g s}{\omega_0} a-h.c.\right)} \left | 0 \right\rangle.
	\label{eq:ad2}
\end{equation}
Here $\left | 0 \right\rangle$ is the bosonic vacuum state and $s=\left \langle \varphi \left| \sigma_z \right| \varphi \right \rangle$. At this stage, the bosonic state $\left | f \right\rangle $ can be used to obtain an effective Hamiltonian, $H_{eff}$, for the spin. The average value of $H$ on the state $\left | f \right\rangle$ provides:
\begin{equation}
	H_{eff}=-\frac{\Delta}{2} \sigma_x + \frac{s^2 g^2}{\omega_0}-2 \frac{s g^2}{\omega_0} \sigma_z,
	\nonumber
\end{equation}
i.e. the adiabatic approximation leads to an effective problem describing a two level system with gap $\Delta$, along $x$ axis, in the presence of a self-consistent magnetic field along $z$ direction. It is straightforward to show that, defining the dimensionless parameter $\lambda=\frac{g^2}{\omega_0 \Delta}$, a self-trapping transition occurs at $\lambda=\frac{1}{4}$. Furthermore $\langle \sigma_z \rangle =0$ for $\lambda < \frac{1}{4}$, whereas $\langle \sigma_z \rangle \ne 0$ for $\lambda \ge \frac{1}{4}$, with $\langle \sigma_z \rangle \propto \sqrt{\lambda-\frac{1}{4}}$, as expected in a mean field transition. Then $M^2=\langle \sigma_z \rangle ^2$ (being $T=0$), differently from BKT quantum phase transition, does not exibit any discontinuity at $\lambda_c=\frac{1}{4}$. Analogously the quantity $X^2=\frac{2 \Delta \lambda}{k} M^2$ increases linearly with $(\lambda-\frac{1}{4})$ near the transition point. Finally the calculation of the average number of phonons provides:
\begin{equation}
	\langle a^{\dagger}a \rangle= \frac{\Delta \lambda}{\omega_0} M^2. 
	\label{avph}
\end{equation}
It vanishes for $\lambda < \frac{1}{4}$ and diverges at the critical point. 

Now we want to address the following question: what happens in the presence of the environment? To this aim we come back to Eq.(\ref{eq:eqSsupp}) where, being interested in the limit $\omega_0 \rightarrow 0$, we perform the substitution: $y=\frac{\omega}{\omega_0}$. The interaction between two classical spins, located at $\tau$ and $\tau^{\prime}$, is given by:
\begin{equation}
	K_{eff}(\tau-\tau^{\prime})=\omega_0 \int\limits_0^{\infty}\!dy J_{eff}(y \omega_0) \frac{ \cosh \left[ y \omega_0 \left( \frac{\beta}{2}-(|\tau-\tau^{\prime}|) \right)\right]}{\sinh \left( \frac{\beta y \omega_0}{2} \right)}.
	\label{uren}
\end{equation}
When $\omega_0 \rightarrow 0$ and $\beta \rightarrow \infty$ with $\omega_0 \beta \rightarrow 0$, Eq.(\ref{uren}) provides:
\begin{equation}
	K_{eff}(\tau-\tau^{\prime}) \rightarrow \frac{2 g^2}{\beta \omega_0}, 
	\label{kom0}
\end{equation}
since
\begin{equation}
	2 \omega_0^3 \alpha_{cav} \int_0^{\infty} \frac{1}{(\omega^2-\omega_0^2)^2+(\pi \alpha_{cav} \omega_0 \omega)^2} d\omega=1.
	\nonumber
\end{equation}
On the other hand Eq.(\ref{kom0}) is independent of $\alpha_{cav}$, i.e. $K_{eff}(\tau-\tau^{\prime})=\tilde{K}(\tau-\tau^{\prime})$, under the assumptions: $\omega_0 \rightarrow 0$ and $\beta \rightarrow \infty$ with $\omega_0 \beta \rightarrow 0$. In other terms, the occurrence of the mean field transition in the adiabatic limit is not influenced by the presence of the environment. This is another difference with BKT quantum phase transition, that is induced by the coupling between the resonator and the bosonic bath.

We end the supplemental material by discussing the behaviour of some quantities of physical interest involving the spin and/or the resonator. 

\begin{figure}[H]
	\centering
	\includegraphics[scale=0.4]{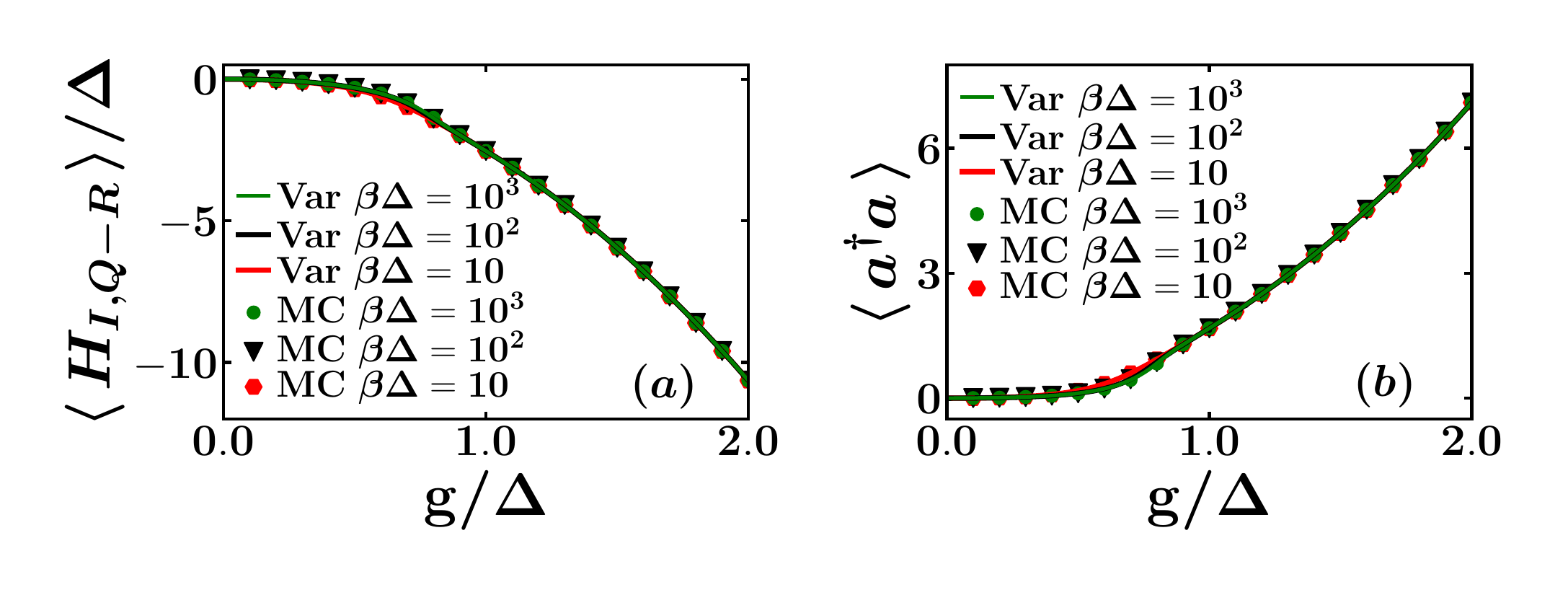}
	\caption{\label{fig:intn} 
		$\langle H_{I,Q-R} \rangle$(a) and $\langle a^{\dagger}a\rangle$(b)(measured with respect to the value in the absence of spin-resonator coupling) as function of $g$ for different temperatures in the variational approach (Var) and WLMC method (MC) ($\omega_0=0.75$ and $\omega_c=10$).}
\end{figure}

In Fig.~\ref{fig:intn}a and Fig.~\ref{fig:intn}b, by using both the variational approach \`a la Feynman and WLMC method, we plot, as function of $g$ and at different temperatures, the average value of the interaction between the spin and the resonator $H_{I,Q-R}= g \sigma_z (a+a^{\dagger})$ and mean number of phonons of the resonator, i.e. $\langle H_{I,Q-R} \rangle$ and $\langle a^{\dagger}a \rangle$ (measured with respect to the value in the absence of spin-resonator coupling). As expected, by increasing the spin-resonator coupling strength $g$, the absolute value of $\langle H_{I,Q-R} \rangle$ increases as well as $\langle a^{\dagger}a \rangle$. In particular, we emphasize that $\langle a^{\dagger}a \rangle$ is finite at $g_c$, whereas in the mean field transition, occurring in the fully adiabatic limit, $\langle a^{\dagger}a \rangle$ diverges.

\begin{figure}[H]
	\centering
	\includegraphics[scale=0.4]{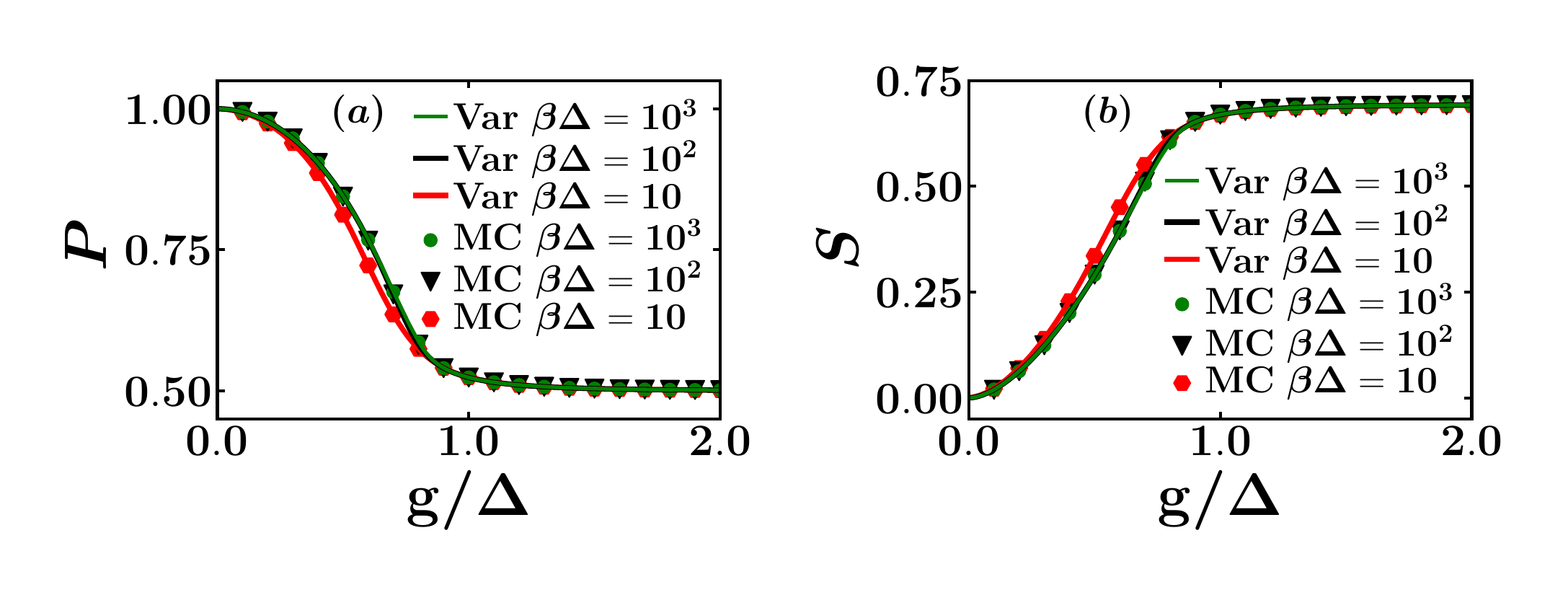}
	\caption{\label{fig:entropy}
		Qubit purity and entropy as function of $g$ for different temperatures in the variational approach (Var) and WLMC method (MC) ($\omega_0=0.75$ and $\omega_c=10$).}
\end{figure}

In Fig.~\ref{fig:entropy}a and Fig.~\ref{fig:entropy}b, by using both the variational approach \`a la Feynman and WLMC method, we plot, as function of $g$ and at different temperatures, the purity and the entropy of the two level system, i.e. $P=Tr(\rho^2)$ and $S=-Tr (\rho \log \rho)$. Here $\rho$ is the qubit density matrix that can be written in terms of the Pauli matrices and vector $\bf {v}$ representing the quantum state on or inside Bloch sphere: $\rho=\frac{1}{2} ( \bf{1} + \bf{v} \cdot \bf{\sigma})$, where $\bf{1}$ is the identity matrix in 2D. On the other hand, for symmetry reasons, $\langle \sigma_y \rangle= \langle \sigma_z \rangle=0$ for any $g $ at finite temperature, and for $g \le g_c$ at $T=0$. Then the two quantities $P$ and $S$ depend only on the value of $\langle \sigma_x \rangle$ ($\bf{v}=$ $(\langle \sigma_x \rangle,0,0)$). In the main text, we have shown that $\langle \sigma_x \rangle$ is $1$ in the absence of spin-resonator coupling and, by increasing $g$, reduces indicating a drop of the effective spin tunneling. Then $P$ ($S$) goes from $1$ ($0$), at $g=0$, to $\frac{1}{2}$ ($\log(2)$) by increasing the strength of the interaction between the qubit and the resonator. In other terms, the two level system: 1) at $g=0$, is in a pure state, represented by a point on the surface of the Bloch sphere; 2) by increasing $g$, $P$ ($S$) reduces (increases) and, for large couplings, $P$ tends to the lowest (highest) possible value, $1/2$ ($\log(2)$), i.e. the qubit state is a completely mixed state, represented by the center of the Bloch sphere.

Finally, differently from the order parameter $M^2$, it is worth mentioning that: 1) all the discussed physical quantities show a weak dependence on the temperature; 2) none of them exhibits a singular behavior at the critical value of the coupling $g_c \simeq 0.9 \Delta$ (it is likely that their derivatives display discontinuities at the quantum phase transition).

\section  {Possible Experimental Implementation.}

\begin{figure}[H]
	\centering
	\includegraphics[scale=0.8]{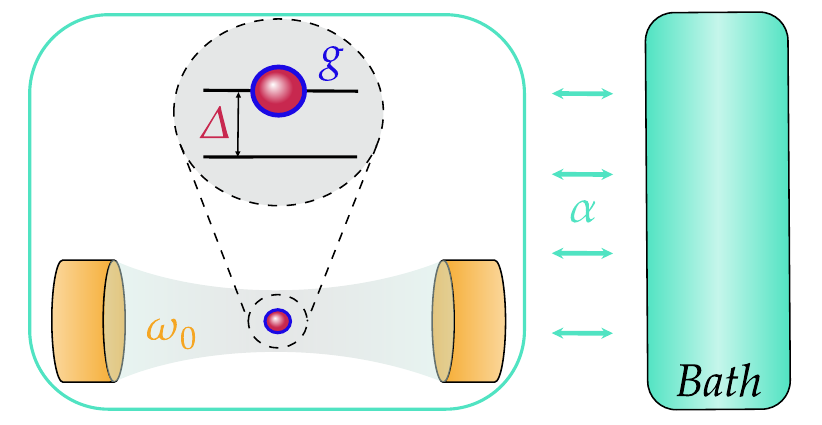}
	\caption{\label{model} 
		Schematic representation of the dissipative Rabi model: a two level system, with gap $\Delta$, interacts with a single mode of a cavity, in turn coupled to an Ohmic bath.}
\end{figure}

\begin{figure}[H]
	\centering
	\includegraphics[scale=0.8]{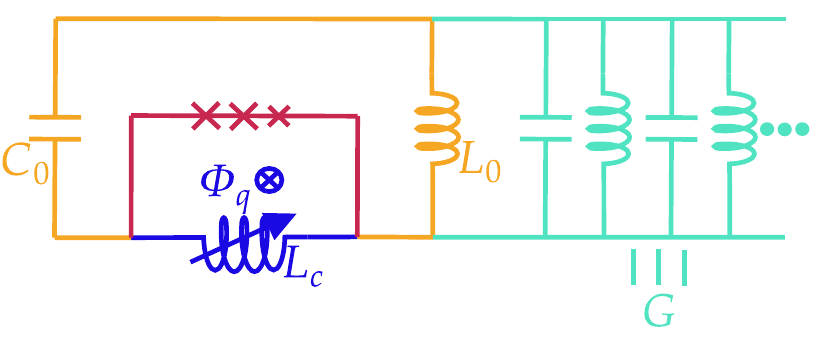}
	\caption{\label{circuit} 
		Circuit diagram. A superconducting flux qubit (red and blue) and a superconducting LC oscillator (blue and yellow) are inductively coupled to each other by sharing a tunable inductance (blue). The dissipative element in the Hamiltonian of Eq.(\ref{eq:definitionH}) is represented by an infinite number of purely reactive elements (green).}
\end{figure}

The dissipative Rabi model, explored in this Letter and sketched in Fig.~\ref{model}, can be implemented by using circuit quantum electrodynamic systems. We start from the experimental realization of the Rabi model, beyond the ultrastrong-coupling regime, got by using superconducting qubit oscillator circuit \citeSupp{naturesupp}. Here a superconducting flux qubit and a superconducting LC oscillator are inductively coupled to each other by sharing a tunable inductance $L_C$. An important feature of the flux qubit is its strong anharmonicity: the two lowest energy levels are well isolated from the higher levels. The corresponding circuit diagram is given in the first part of Fig.~\ref{circuit} (red, blue, and yellow contributions). Our proposal consists in introducing a dissipative element representing the Ohmic bath in our model Hamiltonian (green part of Fig.~\ref{circuit}). Indeed, as specified in the main text, following Devoret \citeSupp{devosupp}, the cavity mode coupled to an Ohmic environment can be represented by a lossy LC circuit, i.e. an LC circuit, in the presence of a dissipative element replaced by an infinite number of purely reactive elements. In the main text we have shown that, for moderate values of the dissipation, the effective resistance that has to be included is of the order of $k\Omega$ and QPT occurs for values of $g_c/\omega_0 \simeq 1$, i.e. $g_c$ lies in the deep strong coupling regime that can be experimentally reached \citeSupp{naturesupp}. 

\bibliographystyleSupp{ieeetr}

\bibliographySupp{suppmat}

\end{document}